\documentclass[useAMS,usenatbib]{pasj02}

\usepackage{comment}
\usepackage{color}
\usepackage{soul}
\sethlcolor{pink}

\usepackage{bm}

\usepackage{graphicx}
\usepackage{lscape}
\usepackage{tikz}
\usepackage{placeins}
\usepackage{cuted}
\usepackage{flushend}
\usepackage[switch,mathlines]{lineno} 

\usepackage{natbib, aas_macros}

\newcommand{\simgt}{\lower.5ex\hbox{$\; \buildrel > \over \sim \;$}}
\newcommand{\simlt}{\lower.5ex\hbox{$\; \buildrel < \over \sim \;$}}

\def\h70kpc{\mathrel{h_{70}^{-1}{\rm kpc}}}

\def\h70Msol{\mathrel{h_{70}^{-1}M_\odot}}

\usepackage{version}

\begin{document}

\Received{}
\Accepted{}



\Received{}
\Accepted{}

\Received{}
\Accepted{}

\title{Constraints on Nonlinearity of the Mass--Concentration Relation\thanks{ Based on data collected at Subaru Telescope, which is operated by the National Astronomical Observatory of Japan.}}

\author{Nobuhiro \textsc{Okabe}\altaffilmark{1,2,3}\orcid{0000-0003-2898-0728}}
\email{okabe@hiroshima-u.ac.jp}
\altaffiltext{1}{Physics Program, Graduate School of Advanced Science and Engineering, Hiroshima University, 1-3-1 Kagamiyama, Higashi-Hiroshima, Hiroshima 739-8526, Japan}
\altaffiltext{2}{Hiroshima Astrophysical Science Center, Hiroshima University, 1-3-1 Kagamiyama, Higashi-Hiroshima, Hiroshima 739-8526, Japan}
\altaffiltext{3}{Core Research for Energetic Universe, Hiroshima University, 1-3-1, Kagamiyama, Higashi-Hiroshima, Hiroshima 739-8526, Japan}


\maketitle

\begin{abstract}
We develop a Bayesian generalization of the previous method \citep{2025A&A...700A..46O} to constrain a wide variety of mass-concentration relations motivated by both theoretical and observational studies. The framework incorporates the intrinsic scatter in concentration, nonlinear weak-lensing (WL) calibration relations, the WL measurement-error covariance, and the underlying parent mass distribution. For computational efficiency, we employ Hamiltonian Monte Carlo method. Our sample consists of 213 X-ray-selected clusters with WL measurements drawn from the eRASS1, XXL, LoCuSS, and MCXC catalogs in the redshift range of $0<z\simlt 0.9$. For each subsample, we account for its specific WL calibration relation and underlying population distribution. We consider a conventional single power-law model, an empirical double power-law model, and four theoretical models calibrated against numerical simulations. The four theoretical mass-concentration models comprise two in which concentration decreases with mass, distinguished by the sign of $d^2\ln c_{200}/d(\ln M_{200})^2$, and two in which the concentration decreases with increasing mass before turning upward at high masses, while $d^2\ln c_{200}/d(\ln M_{200})^2>0$ across the entire mass range.
Because the mass and redshift ranges of our cluster sample do not extend sufficiently into the upturn regime predicted by numerical simulations, we distinguish among the models primarily through the curvature of their mass-concentration relations. Regardless of which model is adopted, the inferred mass-concentration relation, including its redshift evolution, is in good agreement with the results of recent numerical simulations. We find no statistically significant differences among the models at or above the $4\sigma$ level. However, the result shows a tendency toward a gradual flattening of the mass dependence toward higher masses. Specifically, a monotonically decreasing model with positive curvature, $d^2\ln c_{200}/d(\ln M_{200})^2>0$, is preferred over the single power-law relation and a monotonically decreasing model with negative curvature at approximately the $2\sigma$ level and over three double power-law models at approximately the $0.5-1\sigma$ level.
\end{abstract}

\clearpage %


\section{Introduction}

Numerical simulations \citep[e.g.,][]{2001MNRAS.321..559B,2007MNRAS.381.1450N,2013ApJ...766...32B,2014MNRAS.441.3359D,2015MNRAS.452.1217C,2015ApJ...799..108D,2016MNRAS.460.1214L,2016MNRAS.457.4340K,2018ApJ...859...55C,2019ApJ...871..168D,2019MNRAS.486.4001R,
2021MNRAS.506.4210I,2024A&A...681A..67E} indicate the existence of a relationship between halo concentration and halo mass in the NFW model \citep{1996ApJ...462..563N, 1997ApJ...490..493N}. The mass--concentration relation can be broadly categorized into two types.

The first type is the monotonically decreasing model: \citet{2001MNRAS.321..559B} first reported that halo concentration tends to increase toward lower halo masses and lower redshifts. This trend has been confirmed by more recent simulations \citep[e.g.][]{2013ApJ...766...32B,2014MNRAS.441.3359D,2015MNRAS.452.1217C,2018ApJ...859...55C,2019MNRAS.486.4001R,2024A&A...681A..67E}. Recent simulation-based relations \citep[e.g.][]{2013ApJ...766...32B} are formulated in terms of the peak height of the density contrast rather than halo mass. Consequently, their effective slopes may still depend on mass. 
This leads to non-zero second derivative of logarithmic concentration with respect to logarithmic mass, $d^2\ln c/d(\ln M)^2$. They are divided into two categories: the first category is the model \citep[e.g.][]{2013ApJ...766...32B} with a negative second derivative, $d^2\ln c/d(\ln M)^2<0$, and the second category is the model \citep[e.g.][]{2018ApJ...859...55C} with a positive second derivative, $d^2\ln c/d(\ln M)^2>0$. Regardless of the value of the second derivative, the anti-correlation between mass--concentration is naturally interpreted within the context of hierarchical structure formation.
Progenitors of low-mass halos collapse earlier, so their characteristic inner density reflects the higher critical density of the Universe at those earlier epochs. In contrast, more massive halos assemble later, when the background density is lower, and they build up through the accretion and merging of these smaller systems. 

The second type is that the concentration turns upward for the most massive halos \citep{2015ApJ...799..108D,2016MNRAS.460.1214L,2016MNRAS.457.4340K,2019ApJ...871..168D,
2021MNRAS.506.4210I}. The most massive halos are more likely to accrete and capture less massive systems through cluster mergers, resulting in the emergence of the upturn feature in the mass-concentration relation. 
The upturn feature is clearly evident at higher redshifts, $z \simgt 1$, whereas it is much less pronounced in simulation results within the range $0 \le z \le 1$ \citep[e.g.][]{2016MNRAS.457.4340K,2016MNRAS.460.1214L,2019ApJ...871..168D}, even though, in some cases, applying fitting formulas still preserves the upturn feature.

Weak gravitational lensing (WL) provides a unique opportunity to measure cluster mass profiles without requiring assumptions about the dynamical state of the system. However, weak-lensing measurements are sensitive to several sources of systematic uncertainty, including halo triaxiality, the presence of cluster substructures, the choice of cluster center, and the projection of related/unrelated structures along the line of sight. Wide-field optical imaging can cover the full spatial extent of clusters and thereby constrain both their masses and concentrations. Measurements of the cluster mass--concentration relation therefore provide an important test of structure formation on cluster scales \citep[e.g.,][]{2014ApJ...795..163U,2016MNRAS.461.3794O,2016ApJ...821..116U,2017MNRAS.472.1946S,2019PASJ...71...79O,2020ApJ...890..148U,2025A&A...700A..46O,2026arXiv260624142U}. 

However, some early studies were affected by inadequate statistical treatments. These included the omission of the measurement-error covariance between mass--concentration \citep[e.g.,][]{2010PASJ...62..811O} and the neglect of the underlying parent population distribution \citep[e.g.,][]{2016MNRAS.461.3794O}. In regression analyses, ignoring the parent population can bias the inferred slope toward shallower values, a manifestation of regression dilution \citep[e.g.,][]{1996ApJ...470..706A}. Systematic uncertainties in weak-lensing mass measurements must also be taken into account. In particular, survey-quality data typically contain only a finite number density of securely selected background galaxies. Consequently, the accuracy and precision of weak-lensing mass measurements can depend systematically on both cluster redshift and true halo mass. Such effects can propagate directly into the inferred mass-concentration relation and may produce apparent mass or redshift trends if they are not modeled properly. The same can be said of the baseline measurement uncertainties.

\citet{2025A&A...700A..46O} developed a Bayesian framework for the mass--concentration relation that accounts simultaneously for weak-lensing systematics, the measurement-error covariance matrix, and the parent population distribution. Their analysis assumed a linear mass-concentration relation and a linear weak-lensing calibration model. These assumptions have the advantage that the relevant probability distributions can be expressed analytically in Gaussian form, making the calculations computationally efficient and numerically stable. However, such formulations are not sufficiently flexible to describe nonmonotonic or strongly nonlinear features, such as a high-mass upturn in the mass-concentration relation or nonlinear weak-lensing mass--concentration calibrations.

In this paper, we develop a Bayesian generalization of the previous model for jointly modeling nonlinear mass--concentration relations and nonlinear weak-lensing calibrations. The method explicitly incorporates the parent population distribution, covariance matrix of measurement uncertainties, intrinsic scatter, and mass- and redshift-dependent weak-lensing systematics. This generalization allows us to test a broad range of theoretically and empirically motivated mass--concentration models within a common statistical framework.

For this purpose, we compile a sample of 213 X-ray-selected galaxy clusters, as described in Section~\ref{sec:data}. Section~\ref{sec:MCanalysis} presents the Bayesian analysis framework of various mass-concentration models and weak-lensing calibration models. The results and their implications are presented in Section~\ref{sec:result}. Finally, Section~\ref{sec:summary} summarizes our main conclusions.

Throughout this paper, we assume a flat $\Lambda$CDM cosmology with $H_0=100h\,{\rm km\,s^{-1}Mpc^{-1}}$, $\Omega_{m,0}=0.3$ and $\Omega_{\Lambda,0}=0.7$. When computing the spectral index $n$ and peak height $\nu$, we additionally adopt $\sigma_8=0.81$, $n_s=0.9665$, $\Omega_{b,0}=0.05$, and $H_0=70\,{\rm km\,s^{-1}Mpc^{-1}}$

\section{Data} \label{sec:data}

We compiled four X-ray cluster catalogs of the Western Galactic Hemisphere of the eROSITA’s \citep[extended ROentgen Survey with an Imaging Telescope Array;][]{2012arXiv1209.3114M,2021A&A...647A...1P} first All-Sky Survey \citep[eRASS1;][]{2024A&A...685A.106B,2024A&A...688A.210K,2024A&A...689A.298G},  
the Local Substructure ClUster Survey \citep[LoCuSS;][]{2016MNRAS.456L..74S,2016MNRAS.461.3794O}, XXL \citep{2018A&A...620A...5A,2016A&A...592A...1P,2020ApJ...890..148U}, and 
the Meta-Catalogue of X-ray detected Clusters of galaxies \citep[MCXC;][]{2024A&A...688A.187S, 2011A&A...534A.109P}.
X-ray-selected clusters are less affected by projection effects because their emissivity scales with the square of the electron number density, and they also exhibit only a small mis-centering effect \citep[e.g.][]{2023A&A...669A.110O,2025A&A...700A..46O}. The impact of mis-centering in the eRASS1, XXL, and LoCuSS samples has previously been shown to be minimal \citep{2013ApJ...769L..35O,2020ApJ...890..148U,2025A&A...700A..46O}. 
The LoCuSS and MCXC cluster samples are derived from the R\"ontgen SATelleit (ROSAT)
and its associated all-sky survey (RASS) observations, the eRASS1 sample are obtained with the eROSITA instrument onboard the Spectrum Roentgen Gamma (SRG) mission \citep{2021A&A...656A.132S}, and the XXL sample is based on XMM-Newton data, respectively.

We employ two types of weak-lensing shape catalogs. The first is the public HSC-Y3 shape catalog \citep[see][]{2018PASJ...70S..25M,HSC-3Y-Shape}, derived using the re-Gaussianization point-spread-function (PSF) correction technique \citep{2003MNRAS.343..459H} from the Subaru HSC-SSP Survey \citep{2018PASJ...70S...4A,2022PASJ...74..247A} with five bands imaging. The second is a catalog based on a modified implementation \citep{2016MNRAS.461.3794O} of the KSB method \citep{1995ApJ...449..460K}, utilizing Subaru/Suprime-Cam data.

The HSC-SSP data are used for the WL analysis of the eRASS1, XXL, and MCXC clusters. The details are described in \citet{2025A&A...700A..46O}; here we briefly summarize the main properties of the sample selection and analyses. The clusters used for WL mass measurements were selected according to the following criteria: the area fraction within $r<3h_{70}^{-1}\,{\rm Mpc}$, centered on each X-ray centroid, is greater than 70\%, and the innermost radius is smaller than $0.7h_{70}^{-1}\,{\rm Mpc}$, from the parent clusters in the HSC-SSP footprint. Hereafter, we call them WL-targeted clusters.
Background galaxies were selected using the probability threshold $\int_{z_l+0.2}^{\infty} P(z)\,dz > 0.98$ \citep{2018PASJ...70...30M}, where $P(z)$ is the photometric-redshift probability distribution calibrated using spectroscopic data \citep{2018PASJ...70S...9T,2020arXiv200301511N}, and $z_l$ is the cluster redshift. We adopted an adaptive choice of radial bins for the tangential shear profile \citep{2016MNRAS.461.3794O}. If another cluster is present within the maximum radius used to search for the optimal radial binning, this maximum radius is replaced by half of the distance between the two clusters. With a secure background selection using survey-type data, WL mass measurements are still limited by a small number of background galaxies, which can cause individual cluster masses to be accidentally overestimated. The effect should be considered as the WL mass bias (Sec. \ref{subsec:WLbias}) in a study of the mass--concentration relation. 

The Suprime-cam data are used for the WL analysis of the LoCuSS clusters, described in \citet{2016MNRAS.461.3794O}. 
Background galaxies are selected using a radially dependent color--magnitude selection. This selection is constructed by monitoring the consistency between the lensing signal and the average photometric redshift inferred from COSMOS galaxies \citep{2013A&A...556A..55I} that are close to each source galaxy in color--magnitude space. The resulting number densities of background galaxies are about twice that of the HSC-SSP in the same redshift range. The choice of radial bins is the same as described above, except that the presence of surrounding clusters is taken into account. Using the same setup as in the actual analysis, such as the number density distribution of background galaxies, redshift, and mass range, \citet{2016MNRAS.461.3794O} confirmed that the WL measurements recover the input parameters through a toy NFW model and 23 numerically-simulated clusters  \citep{2012MNRAS.421.1073B}, but with only up to $30\times50$ realizations. 

We adopted the same WL mass measurements for the eRASS1 and LoCuSS clusters as those reported by \citet{2025A&A...700A..46O} and \citet{2016MNRAS.461.3794O}, respectively. For the eRASS1 clusters, 78 systems were targeted for WL measurements, including three clusters affected by misassociation, and WL masses were successfully determined for 72 of them.
For the LoCuSS sample, WL masses were measured for 50 clusters in the redshift range $0.15\le z \le 0.3$. Since one of these clusters overlaps with the eRASS1 sample, we adopted the HSC-SSP mass, which is consistent with the LoCuSS mass.

For the XXL and MCXC clusters, we repeated the WL mass measurements using the same method as in \citet{2025A&A...700A..46O}. Although WL masses for the XXL clusters have been reported by \citet{2020ApJ...890..148U}, we performed a homogeneous analysis in order to evaluate the WL mass bias self-consistently. We successfully measured WL masses for 95 of the 138 WL-targeted XXL clusters. We used the second version of the MCXC cluster catalog \citep{2024A&A...688A.187S}. After excluding clusters that overlap with the other X-ray cluster catalogs, we measured WL masses for 12 of the 16 WL-targeted clusters. 
From visual inspection of the HSC images, we identify offsets of approximately $280\,h_{70}^{-1}\,\mathrm{kpc}$ between the BCGs and the X-ray centroids in two of three nearest clusters at $z<0.1$.
Excluding these clusters from the analysis does not affect the results presented in the paper. We thus keep them in the following analysis.

We then exclude any clusters whose angular separation from the nearest neighboring cluster on the sky is smaller than the WL $r_{200}$, in order to minimize lensing contamination from non-target clusters. Here, $r_{200}$ is the overdensity radius within which the mean mass density is 200 times the critical density at the cluster redshift. To identify nearby projected clusters, we use the combined catalog of eRASS1, XXL, LoCuSS, and MCXC clusters. In this process, we remove 2 eRASS1 and 13 XXL clusters from our sample. The final sample used in this paper consists of 213 X-ray-selected clusters.

\section{Bayesian analysis for the mass--concentration relation} \label{sec:MCanalysis}

Building on the earlier study of \citet{2025A&A...700A..46O}, we outline a Bayesian framework for the mass--concentration relation that incorporates WL calibration.

We adopt notation for the logarithms of the true, true WL, and observed WL quantities. For mass, we define
$X = \ln (M_{200}^{\rm true}/M_p)$, $\tilde{x} = \ln (M_{200}^{\rm WL,true}/M_p)$, and $x = \ln (M_{200}^{\rm WL}/M_p)$, where $M_p$ denotes the pivot mass. Similarly, for concentration, we define
$Y = \ln (c_{200}^{\rm true}/c_p)$, $\tilde{y} = \ln (c_{200}^{\rm WL,true}/c_p)$, and $y = \ln (c_{200}^{\rm WL}/c_p)$, where $c_p$ denotes the pivot concentration.

Throughout this paper, we set $c_p=4$ and $M_p=10^{14}h^{-1}M_\odot$. We also use the vector notation $\bm{X}_i = (X_i, Y_i)$, $\tilde{\bm{x}}_i = (\tilde{x}_i, \tilde{y}_i)$, and $\bm{x}_i = (x_i, y_i)$ to denote the corresponding variables for the $i$-th cluster.

When the $i$-th cluster belongs to the $a$-th subsample, the hierarchical Bayesian model for the mass--concentration relation is written as
\begin{align}
  &  p(\bm{x} \mid z , \bm{\theta},\bm{\theta}_{\rm cosmo},\bm{\theta}_{\rm WL}, \bm{\psi}) \label{eq:likelihood} \\
\    &\propto\prod_a^{N_a}
    \prod_{i\in a}^N \int d^2\tilde{\bm{x}}_i \int dX_i \int dY_i p(\bm{x}_i \mid\tilde{\bm{x}}_i)p(\tilde{\bm{x}}_i \mid X_i,Y_i,z_i, {\bm \theta}_{{\rm WL},a}) \nonumber\\
    &\times  p(Y_i  \mid X_i, z_i, {\bm \theta}, \bm{\theta}_{\rm cosmo})p(X_i \mid z_i,{\bm \psi}_a)\nonumber \\
    &\times p(\bm{\theta})p(\bm{\theta}_{\rm cosmo})p(\bm{\theta}_{{\rm WL},a})p(\bm \psi_a) \nonumber 
\end{align}
Here, $z_i$ denotes the redshift of the $i$-th cluster, $\bm{\theta}$ represents the parameters of the mass--concentration relation, $\bm{\theta}_{\rm cosmo}$ denotes the cosmological parameters, and $\bm{\theta}_{\rm WL}$ represents the weak-lensing bias parameters, respectively. The parameters $\bm{\psi}_a$ describe the parent mass distribution of the $a$-th subsample. Each conditional probability in eq. \ref{eq:likelihood} is interpreted in the following subsections. The double numerical integrations along $X$ and $Y$ are computed by importance sampling (Appendix \ref{app:white}).

\subsection{Measurement Error}

The observed WL mass and the corresponding true WL mass are related through the error covariance matrix, $\bm{\Sigma}_{\rm err}$. Specifically, we use
\begin{align}
    p(\bm{x}_i \mid \tilde{\bm{x}}_i)
    = \mathcal{N}\!\left(\bm{x}_i \mid \tilde{\bm{x}}_i, \bm{\Sigma}_{{\rm err},i}\right),
\end{align}
where $\mathcal{N}(\bm{x} \mid \bm{\mu}, \bm{\Sigma})$ denotes a multivariate normal distribution with mean vector $\bm{\mu}$ and covariance matrix $\bm{\Sigma}$.

\subsection{WL mass bias} \label{subsec:WLbias}

The conditional probability, $p(\tilde{\bm{x}}_i \mid X_i,Y_i,z_i, {\bm \theta}_{\rm WL})$, is described by the relationship between the true WL masses and the true masses, namely, WL mass bias. In general, the WL mass bias consists of two effects: astrophysical (intrinsic) properties and observational conditions. The former is mainly due to the orientation of elliptical halos, the presence of subhalos, and their surroundings \citep[e.g.][]{2010A&A...519A..90M,2011ApJ...740...25B,2012MNRAS.426.1558G,2024A&A...681A..67E}. The latter effect corresponds to a positive bias in the measured WL mass, arising from the limited number of background galaxies, which dominate in the survey data, as noted by \citet{2025A&A...700A..46O}. 

In this paper, we take the latter effect into account for two reasons. First, the adaptive-choice binning scheme helps suppress the former effect \citep{2016MNRAS.461.3794O}.  
Second, assessing the WL calibration requires a data set that is substantially larger than the sample used in this study.
Since the number and redshift coverage of clusters in the numerical simulations are limited, the latter effect can be explicitly quantified from a toy model. To assess the WL calibration, we adopted a single NFW model whose logarithmic mass, $\ln M_{200}$, is uniformly distributed over the range $[\ln (0.8 \times 10^{14}\,h_{70}^{-1}M_\odot), \ln (30 \times 10^{14}\,h_{70}^{-1}M_\odot))$, and the same background densities in the real observations. 
Our sample comprises two distinct datasets, one derived from the HSC-SSP Survey and the other from Suprime-Cam data for the LoCuSS survey \citep{2016MNRAS.461.3794O}, which differ substantially in both the background-galaxy selection methodology and the resulting background-galaxy number densities.

\subsubsection{HSC-SSP}

\begin{figure}
    \includegraphics[width=\linewidth]{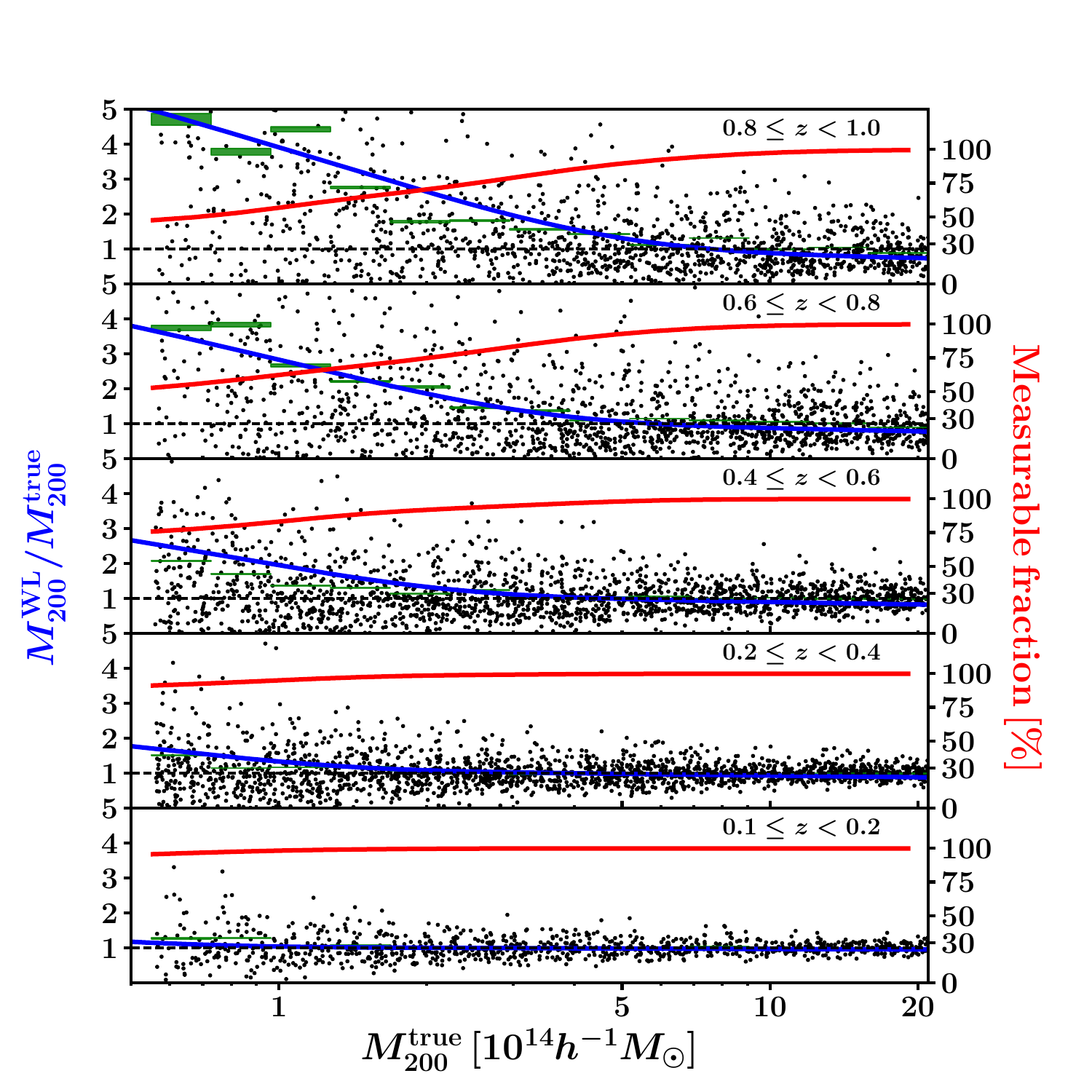}
    \caption{Mass bias, $M^{\rm WL}_{200}/M_{200}^{\rm true}$, in five redshift bins. The black points are the mass bias for individual mock clusters. We ignore their errors for visual purposes. The green boxes represent the WL bias evaluated via the noise-weighted geometric mean within each mass bin. The red solid lines are the measurable fraction. The blue solid lines are the best-fit scaling relations of eq. \eqref{eq:WLmassCalib}. {Alt text: one figure containing five vertical panels.}}
    \label{fig:mbias_vs_M_WLcal}
\end{figure}

\begin{figure}
    \includegraphics[width=\linewidth]{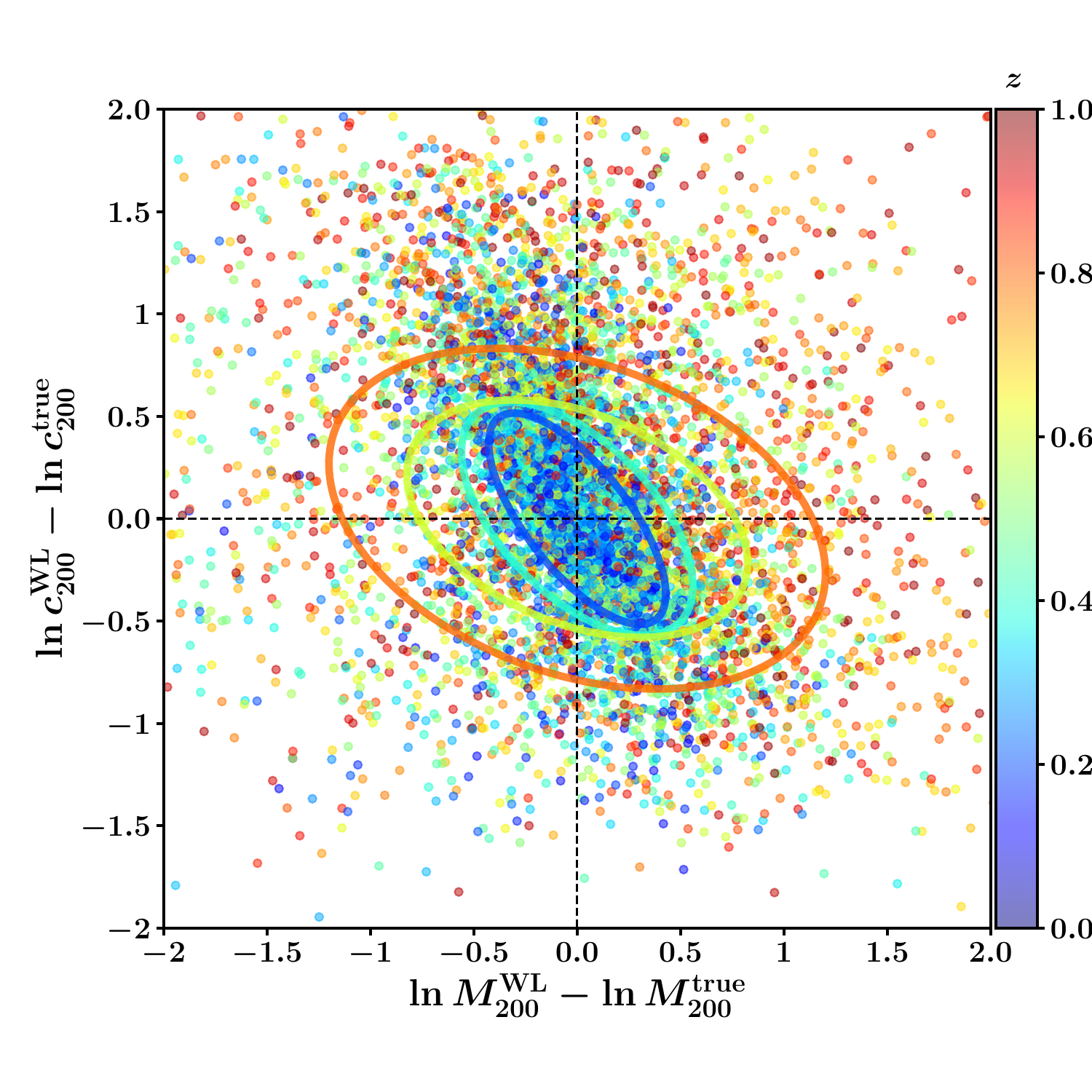}
    \caption{The points indicate the logarithmic WL biases, with their colors corresponding to different redshifts. The blue, green, light-green, and orange ellipses show the precision-weighted measurement covariance matrix at z = 0.2, 0.4, 0.6, and 0.8, respectively. {Alttext: one figure in color.}}
    \label{fig:logbias}
\end{figure}

We use the results from \citet{2025A&A...700A..46O} using 9,000 mock clusters. \citet{2025A&A...700A..46O} found that the average WL mass bias exhibits a strong correlation with the measurable fraction. Within the mass range of the eRASS1 sample, the measurable fraction is high, and the WL calibration remains approximately linear. They therefore adopted a calibration based on a linear function. This allows us to reduce the computational cost by using the Gaussian chain rule. However, toward lower masses and higher redshifts, the mass bias becomes increasingly nonlinear. It is therefore important to explicitly account for the nonlinear nature of these effects when deriving the mass--concentration relation, because otherwise the inferred relation may be biased. This is also an essential step in preparation for Stage-IV surveys. In the present work, we introduce the following expression, which remains applicable over the entire mass range.

\begin{align}
 p(\tilde{\bm{x}} \mid X,Y,z, {\bm \theta}_{\rm WL})
 &=\mathcal{N}\Bigl(\tilde{\bm{x}} \;\Big|\; 
 {\bm f}_{\rm WL}
\begin{pmatrix} X \\ Y \end{pmatrix}, \; {\bm \Sigma}_{\rm WL}(z) \Bigr) \\
{\bm f}_{\rm WL}\begin{pmatrix} X \\ Y \end{pmatrix}&=\bm{\alpha}'_{\rm WL} + \bm{\beta}'_{\rm WL}\begin{pmatrix} X \\ Y \end{pmatrix}+\nonumber \\
&\begin{pmatrix}
   \zeta_{\rm WL}' \dfrac{1}{1+\exp(\exp X/\exp \tau_{\rm WL}')} \\ 0
\end{pmatrix} \label{eq:WLmassCalib} \\
\bm{\alpha}_{\rm WL}'(z) &= \bm{\alpha}_{\rm WL}+\bm{\gamma}_{\rm WL}F(z), \nonumber \\
     \bm{\beta}_{\rm WL}'(z) &= \bm{\beta}_{\rm WL}+\bm{\delta}_{\rm WL}F(z), \nonumber \\
\zeta_{{\rm WL},X}'&= \zeta_{{\rm WL},X} +  \xi_{{\rm WL},X}F(z), \nonumber  \\
\ln \tau_{{\rm WL},X}'&= \ln \tau_{{\rm WL},X}(1  +\nu_{{\rm WL},X}F(z)) , \nonumber\\
F(z)
&=
\ln\left(\frac{1+z}{1+z_p}\right). \nonumber
\end{align}
Here, $\zeta_{\rm WL}$ characterizes the amplitude of the non-linear deviation term. Recall that $X$ is defined as a logarithmic quantity. Therefore, $\exp (X)$ in the deviation term directly corresponds to the true WL mass. The best-fit line agrees well with the WL bias across the redshift-mass bins (Figure \ref{fig:mbias_vs_M_WLcal}). The logarithmic WL biases are displayed in Figure \ref{fig:logbias}. The characteristic measurement uncertainties are depicted as color ellipses. We computed  a precision-weighted measurement covariance matrix, $\sum_i^N({\bm \Sigma}_{{\rm err},i}^{-1}/N)^{-1}$. As the redshift increases, the distribution becomes progressively more elongated along the $\ln M$ direction than the $\ln c$ direction. The best-fit parameters are shown in Table \ref{tab:WLcalib}. We employ the pivot redshift, $z_p=0$.

The redshift-dependent intrinsic covariance is specified by
\begin{align}
\bm{\Sigma}_{\rm WL}(z)&=\Bigl(\begin{smallmatrix}
       \sigma_{{\rm WL},X}^2(z) &  r_{{\rm WL},XY}\sigma_{{\rm WL},X}(z)\sigma_{{\rm WL},Y}(z)  \\
       r_{{\rm WL},XY}\sigma_{{\rm WL},X}(z)\sigma_{{\rm WL},Y}(z) & \sigma_{{\rm WL},Y}^2(z) 
    \end{smallmatrix}\Bigr), \nonumber \\
    \sigma_{\rm WL}(z)&=\exp(\ln \sigma_{\rm WL}+\gamma_{\sigma_{\rm WL}}F(z)), \nonumber \\
    r_{XY}&=\tanh u_{r_{XY}}.\nonumber 
\end{align}
We reparameterize the correlation coefficient using $u_{r_{XY}}$ to ensure that it always remains within the physical range $-1<r_{XY}<1$.

\begin{table}[]
    \centering
    \begin{tabular}{c|cc}
    & HSC-SSP & LoCuSS \\ 
    \hline
    $\alpha_{{\rm WL},X}$ & $-0.028_{-0.056}^{+0.058}$	 & $-0.037_{-0.098}^{+0.101}$  \\
    $\beta_{{\rm WL},X}$ & $1.008_{-0.022}^{+0.022}$	 & $1.010_{-0.044}^{+0.043}$\\
    $\gamma_{{\rm WL},X}$ & $0.491_{-0.265}^{+0.262}$	 & $0.695_{-0.499}^{+0.488}$ \\
    $\delta_{{\rm WL},X}$ & $-0.316_{-0.101}^{+0.098}$	 & $-0.308_{-0.213}^{+0.219}$ \\
      $\ln \tau_{{\rm WL},X}$ & $-1.767_{-0.547}^{+0.497}$ & - \\ 
  $\nu_{{\rm WL},X}$ & $-2.326_{-0.299}^{+0.221}$	 & - \\
$\zeta_{{\rm WL},X}$ & $0.724_{-0.898}^{+1.255}$	  & 0 \\
$\xi_{{\rm WL},X}$ & $3.113_{-2.313}^{+1.873}$	& 0 \\
$\alpha_{{\rm WL},Y}$ & $-0.017_{-0.034}^{+0.032}$	 &  $0.004_{-0.082}^{+0.083}$ \\
$\beta_{{\rm WL},Y}$ & $1.113_{-0.052}^{+0.053}$	 &  $1.119_{-0.158}^{+0.156}$ \\
$\gamma_{{\rm WL},Y}$ & $0.742_{-0.107}^{+0.118}$	 &  $0.301_{-0.408}^{+0.412}$ \\
$\delta_{{\rm WL},Y}$ & $-0.892_{-0.191}^{+0.177}$ &   $-0.799_{-0.778}^{+0.783}$ \\
$\ln \sigma_{{\rm WL},X}$  &$-1.807_{-0.087}^{+0.084}$	 &  $-2.326_{-0.082}^{+0.076}$ \\
$\gamma_{\sigma{\rm WL},X}$ & $1.016_{-0.249}^{+0.236}$	 & 0 \\
$\ln \sigma_{{\rm WL},Y}$  &$-1.291_{-0.053}^{+0.052}$	&   $-1.582_{-0.039}^{+0.040}$\\
$\gamma_{\sigma{\rm WL},Y}$ &$1.273_{-0.141}^{+0.151}$ & 0 \\
$u_{XY}$ & $-4.030_{-0.663}^{+0.487}$	& $-3.292_{-0.665}^{+0.506}$
    \end{tabular}
    \flushleft
    \caption{WL calibration parameters for the HSC-SSP and the LoCuSS datasets.}
    \label{tab:WLcalib}
\end{table}

\subsubsection{LoCuSS}

\citet{2016MNRAS.461.3794O} employed a radius-dependent magnitude-color selection using deep Suprime-Cam imaging for clusters at $0.15 \le z \le 0.3$. In this redshift interval, the background galaxy number density is more than twice that of the HSC-SSP. Following \citet{2016MNRAS.461.3794O}, we implemented a radially varying background distribution for 3,000 mock clusters. Over this redshift and mass range, the WL bias can be described by a linear relation. Accordingly, we set $\zeta_{{\rm WL},X}=\xi_{{\rm WL},X}=\gamma_{\sigma{\rm WL},X}=\gamma_{\sigma{\rm WL},Y}=0$ (Table \ref{tab:WLcalib}).

\subsection{mass--concentration relations}

In our model, the logarithmic concentration $Y$ is not assumed to be uniquely determined by the logarithmic mass $X$ and redshift $z$, but is instead allowed to scatter around a mean relation. We describe this conditional distribution as
\begin{align}
     p(Y \mid X, z, {\bm \theta}, \bm{\theta}_{\rm cosmo})
     =
     {\mathcal N}\!\left(
     Y \mid \mu(X,z, \bm{\theta}, \bm{\theta}_{\rm cosmo}), \sigma_Y
     \right),
\end{align}
where $\mu(X,z,\bm{\theta},\bm{\theta}_{\rm cosmo})$ is the model prediction for the mean logarithmic concentration, and $\sigma_Y$ is the intrinsic scatter. We treat $\ln \sigma_Y$ as a parameter of intrinsic scatter. This Gaussian formulation provides a simple and flexible probabilistic description of the mass--concentration relation, incorporating both the mean trend and the intrinsic scatter around it. We consider the following six models for the mass--concentration relation.

\subsubsection{Single-power law (SP)} \label{subsec:SP}

We first consider a standard empirical relation between halo mass--concentration \citep[e.g.][]{2001MNRAS.321..559B}, parameterized by a single power (SP) law:
\begin{align}
    c
    &=
    c_0 \left(\frac{1+z}{1+z_p}\right)^{\gamma}
    \left(\frac{M}{M_p}\right)^\beta,
\end{align}
where $c_0$ is the normalization at the pivot redshift $z_p=0$ and $\beta$ and $\gamma$ represent the slope associated with the mass dependence and the slope that captures the redshift dependence of the normalization, respectively.
In logarithmic notation, this relationship can be expressed as
\begin{eqnarray}
    \mu = \ln \frac{c}{c_p}=\alpha + \beta X + \gamma F(z)-\ln c_p. \label{eq:SP}
\end{eqnarray}
In the context of hierarchical structure formation, the slopes describing the mass dependence and redshift dependence are theoretically expected to satisfy $\beta<0$ and $\gamma<0$. We therefore reparameterize them using $\xi_\beta$ and $\xi_\gamma$ as
\begin{align}
    \beta&=-\ln(1+e^{\xi_\beta}), \label{eq:xi_1} \\   
    \gamma&=-\ln(1+e^{\xi_\gamma}). \label{eq:xi_2}
\end{align}
In the limit $\xi\rightarrow-\infty$, both $\beta$ and $\gamma$ approach zero. This reparameterization enables a fair comparison with results based on nonlinear mass-concentration relations in which the slopes are constrained to be either positive or negative.
Here, ${\bm \theta}=\{\alpha=\ln c_0, \xi_\beta, \xi_\gamma\}$ are the free parameters.

\subsubsection{Bhattacharya et al. (2013) model (B13) } \label{subsec:B13}

\citet[][hereafter B13]{2013ApJ...766...32B} calibrated the halo mass-concentration relation using cosmological gravity-only $N$-body simulations under the WMAP7 cosmology \citep[][$\sigma_8=0.8$]{2011ApJS..192...18K}, covering $M_{200}\sim 2\times10^{13}-10^{14}h^{-1}M_\odot$ in a redshift range of $z=0-2$. The simulation data shows that the mass concentration decreases monotonically as mass increases. 
They demonstrated that concentration is a function of the peak height, $\nu$,  and a linear growth factor of density fluctuations normalized to unity at $z=0$, $D(z)$, given by
\begin{align}
   c(\nu) &= c_0 D(z)^\gamma \nu^{\beta}, \\
    \nu(M,z)&= \frac{1}{D(z)}\left(1.12\left(\frac{M}{5\times 10^{13}h^{-1}M_\odot}\right)^{0.3}+0.53\right),  \nonumber \\
    D_+(z)&=\frac{5}{2}\Omega_{m,0}E(z) \int^\infty_z \frac{(1+z)}{E^3(z)}dz, \label{eq:linearGrowth0}\\
    D(z)&=D_+(z)/D_+(0), \label{eq:linearGrowth} \\
    E(z)&=\sqrt{\Omega_{m,0}(1+z)^3+\Omega_{\Lambda,0}} \nonumber 
\end{align}
finding that the slope of the $c_\Delta$-$\nu$ relation remains essentially unchanged with redshift, while its normalization shifts by about $\sim 30$\% at the high-mass end. The formula is not universal and varies by about $\sim 20$ \% when the cosmological parameters are modified.
The logarithmic form of the relation can be written as
\begin{align}
\mu =\alpha + \beta \ln \nu(X,z) + \gamma \ln D(z) - \ln c_p, \label{eq:B13}
\end{align}
where $\alpha=\ln(c_0)$, $\beta$, and the combination of $\gamma$ with $\beta$ characterize the overall normalization, the mass-dependent slope, and the redshift-dependent slope, respectively. Given $\sigma^2 \propto M^{-(n_{\rm eff}+3)/3}$ where $n_{\rm eff}$ is an effective spectral index, a second derivative of the baseline becomes $d^2\ln \mu/d(\ln M)^2 \propto \beta (d n_{\rm eff}/d\ln M)/6$. As halo mass increases, $n_{\rm eff}$ becomes progressively less negative due to the scale-dependent suppression of small-scale power in the CDM power spectrum.
Thus, $dn_{\rm eff}/d\ln M>0$. Since $\beta<0$, this leads to leads to $d^2 \ln \mu/d(\ln M)^2 <0$. The negative curvature indicates that its logarithmic slope becomes progressively more negative toward higher masses. 

For all halos, the B13 parameters corresponding to $M_{200}$ are $\alpha = \ln(5.9)$, $\beta = -0.35$, and $\gamma = 0.54$, respectively, and the intrinsic scatter is $\sigma_{\rm ln c_\Delta} = 0.33$. By analogy with Sec. \ref{subsec:SP}, we define the $\xi$-parameters via $\beta=-\ln(1+e^{\xi_\beta})$ and $\gamma=\ln(1+e^{\xi_\gamma})$. In other words, the set of free parameters is given by ${\bm \theta}=\{\alpha, \xi_\beta, \xi_\gamma\}$.

\subsubsection{Child et al. (2018) model (C18)} \label{subsec:C18}

\citet[][hereafter, C18]{2018ApJ...859...55C} run gravity-only $N$-body simulations down to a minimum halo mass of $M_{200}\sim 2 \times 10^{11}M_\odot$ covering the redshift range of $z=0-4$. Their simulation results indicate that mass concentration declines steadily with increasing mass, but this dependence on mass weakens at higher redshifts. To describe such a trend, they introduce a functional form, specified by 
\begin{align}
    c=A\left[\left(\frac{M/M_*}{b}\right)^\beta \left(1+ \frac{M/M_*}{b}\right)^{-\beta} -1 \right ] +c_0 \label{eq:C18A}
\end{align}
Here, $M_*$ is a redshift-dependent nonlinear mass scale given by $\sigma(R_*,z)=\delta_c$ with the amplitude of mass fluctuation smoothed with a top-hat filter, $\sigma(M,z)$ and $\delta_c=1.686$. The parameter $b$ characterizes the scale at which the transition occurs.
Their best-fit parameters are $A=3.44, b=430.49, \beta=-0.10$ and $c_0=3.19$ for all individual halos.  

This functional expression (eq. \ref{eq:C18A}) becomes numerically unstable depending on the chosen sign of $\beta$. We therefore recast the baseline in the form
\begin{align}
\mu &= \ln \frac{A}{c_p} \label{eq:C18} \\
& \qquad + \ln \left[\left(\frac{1}{1+\exp\left(-(X-X_*-\ln b\right))}\right)^{\beta}-1+\exp(\ln c_r)\right], \nonumber 
\end{align}
where $c_r=c_0/A$, $X_*=\ln M_*/M_p$, and $\beta=-\ln(1+e^{\xi_\beta})$. 
The free parameters are ${\bm \theta}=\{\ln A, \ln b, \ln c_r, \xi_\beta\}$.
Because the logarithmic term includes a sigmoid function, $S({X - X_+ - \ln b})$, the concentration tends to a flat, asymptotic value as the mass increases, which is a characteristic property of this model. Thus, the model has a positive second derivative, $d^2 \mu/d(\ln M)^2>0$.

\subsubsection{Double-power law (DP)} \label{subsec:DP}

As motivated by recent numerical simulations \cite[e.g.][]{2015ApJ...799..108D,2016MNRAS.460.1214L,2016MNRAS.457.4340K}, we extend the conventional single-power-law mass--concentration relation to a double-power-law form in order to allow for a possible upturn at the high-mass end. The model is written as
\begin{align}
c
&=
c_0
\left(\frac{M}{M_p}\right)^{\beta}
\left(\frac{1+z}{1+z_p}\right)^{\gamma}
\left[
1+
\left(
\frac{M}{M_t(z)}
\right)^{\beta_2-\beta}
\right].
\end{align}
The second term inside the brackets adds another component that can account for deviations from a simple power-law behavior, where $\beta_2>0$ expresses the upturn power-law slope. The characteristic mass at which the slope changes is denoted by $M_t(z)$. We adopt the following functional form:
\begin{align}
M_t(z)
=
M_p e^{\alpha_{t}}\left(\frac{1+z}{1+z_p}\right)^{\gamma_t}.
\end{align}
with $z_p=0$. Using logarithmic variables, the above equation can be rewritten as
\begin{align}
\mu &=\alpha +\beta X+\gamma F(z)\nonumber \\ 
&\quad  +\ln\left[1+\exp\left((\beta_2-\beta)(X-X_t(z))\right)\right]-\ln c_p, \label{eq:DP}  \\
X_t(z)&=\alpha_t+\gamma_{t}F(z). \nonumber 
\end{align}
In comparison with eq.~\ref{eq:SP}, the fourth term accounts for the upturn component.  
This parameterization reduces degeneracies between parameters and facilitates tighter constraints on them.
In the Bayesian analysis, we introduce the parameters $\xi$ via the relations $\beta=-\ln(1+e^{\xi_\beta})$, $\beta_2=\ln(1+e^{\xi_{\beta_2}})$, and $\gamma=-\ln(1+e^{\xi_{\gamma}})$, in order to satisfy the required conditions. We set $\gamma_t=0$ because it cannot be constrained. 
The parameter set is ${\bm \theta}=\{\alpha, \xi_\beta, \xi_{\gamma}, \xi_{\beta_2}, \alpha_t\}$.

\subsubsection{Diemer \& Kravtsov (2015) model (DK15)}

\citet[][hereafter, DK15]{2015ApJ...799..108D} studied the mass--concentration relation using different types of dissipationless $\Lambda$CDM simulations
with WMAP7 and Planck cosmologies, and self-similar models with different power-law matter power spectra. 
DK15 proposed a unversal concentration model in which the halo concentration is described as a function of the peak height $\nu$ and the local slope of the linear matter power spectrum. Following \citet{2015ApJ...799..108D}, we write
\begin{align}
    \mu &=\ln \frac{c}{c_p}=
    \ln\frac{c_{\rm min}}{2c_p}
    +\ln\left[
    \left(\frac{\nu}{\nu_{\rm min}}\right)^{\beta}
    +
    \left(\frac{\nu}{\nu_{\rm min}}\right)^{\beta_2}
    \right],
    \label{eq:DK15}
    \\
    c_{\rm min}
    &=
    \phi_0+\phi_1 n,
    \\
    \nu_{\rm min}
    &=
    \eta_0+\eta_1 n.
\end{align}
The peak height is defined by $\nu=\delta_c/(\sigma(M,z=0)D(z))$ with $\delta_c=1.686$ and the linear growth factor of eq \ref{eq:linearGrowth}.
Here, $n$ is the local slope of the linear matter power spectrum, defined as
\begin{align}
    n(k)=\frac{d\ln P(k)}{d\ln k}\Bigr|_{k=\kappa 2\pi/R}.
\end{align}
Here, $P(k)$ denotes the linear matter power spectrum. We adopt the no-wiggle transfer function of \citet{1998ApJ...496..605E}, i.e., the Eisenstein \& Hu model without baryon acoustic oscillations. The effective slope is evaluated at the wavenumber
\begin{align}
    k_R=\kappa \frac{2\pi}{R},
\end{align}
where $R$ is the Lagrangian radius of the halo. Throughout this work, we fix $\kappa=1$, following the updated parameters of \citet{2019ApJ...871..168D}. The updated median parameters are $\phi_0=6.58,\phi_1=1.27,\eta_0=7.28,\eta_1=1.56,\beta=1.08$, and $\beta_2=1.77$, respectively. This model has a convex concentration-mass relation in log-log space; therefore, $d^2\ln c/d(\ln M)^2>0$.
We note, however, that the simulated data at $z=0$ do not exhibit any indication of a upturn in concentration, even up to the maximum halo mass shown in Figure 5 of \citet{2015ApJ...799..108D}.

For numerical stability, we replace the original parameters $\phi_0, \phi_r, \eta_0$, and $\eta_r$ with their logarithms $\ln \phi_0, \ln \phi_r, \ln \eta_0$, and $\ln \eta_r$, respectively, and rewrite the expressions as
\begin{align}
    \ln c_{\rm min}&=\ln \phi_0 + \ln (1 + \ln \phi_r n), \\
    \ln \nu_{\rm min}&=\ln \eta_0 + \ln (1 + \ln \eta_r n). 
\end{align}
As with other models, we introduce parameters $\xi$ to describe the negative and positive slopes, with $\beta=-\ln(1+e^{\xi_\beta})$ and $\beta_2=\ln(1+e^{\xi_{\beta_2}})$. Thus, the full parameter vector is ${\bm \theta}=\{\xi_\beta, \xi_{\beta_2}, \ln \phi_0, \ln \phi_r, \ln \eta_0, \ln \eta_r, \kappa\}$, which specifies the baseline concentration via the intermediate quantities $n$ and $\nu$. These are functions of halo mass, redshift, and the chosen cosmology.

\subsubsection{Diemer \& Joyce (2019) model (DJ19) and Ishiyama et al. (2021) model (I21)}

\citet[][hereafter, DJ19]{2019ApJ...871..168D} developed an analytical framework for the evolution of the halo concentration-mass relation based on pseudo-evolution, and extend it phenomenologically using the effective exponent of linear growth, $\alpha_{\rm eff}$ and the effective slope of the power spectrum, $n_{\rm eff}$, to describe halos over all masses. \citet[][hereafter, I21]{2021MNRAS.506.4210I} also applied this formulation. The formula is represented by 
\begin{align}
    c=C(\alpha_{\rm eff}) \tilde{G}\left(\frac{A(n_{\rm eff})}{\nu}\left[1+\frac{\nu^2}{B(n_{\rm eff})}\right]\right)
\end{align}
where ${\tilde G}$ is an inverse function of 
\begin{align}
    G(x)&=\frac{x}{(g(x))^{(5+n_{\rm eff})/6}}, \\
    g(x)&=\ln(1+x)-\frac{1}{1+x}. 
\end{align}
The normalization parameters are specified as 
\begin{align}
    A(n_{\rm eff})&=a_0 (1+ a_1(n_{\rm eff}+3)), \nonumber \\
    B(n_{\rm eff})&=b_0 (1+ b_1(n_{\rm eff}+3)), \nonumber \\
    C(n_{\rm eff})&=1- c_\alpha (1- \alpha_{\rm eff}), \label{eq:DJ19} \\
    \alpha_{\rm eff}(z)&= -\frac{d \ln D(z)}{d \ln(1+z)}, \nonumber \\
    n_{\rm eff}(M)&=-2 \frac{d \ln \sigma(R)}{d \ln R}\Bigr|_{R=\kappa R_L}-3,
\end{align}
where $D(z)$ is the linear growth factor (eq. \ref{eq:linearGrowth}) and $R_L$ is the Lagrangian radius of a halo. The median parameters of numerical simulations \citep{2019ApJ...871..168D} are $\kappa=0.41,a_0=2.45,a_1=1.82,b_0=3.20,b_1=2.30$, and $c_\alpha=0.21$, respectively. This model has a convex concentration-mass relation in log-log space; therefore, $d^2\ln c/d(\ln M)^2>0$.

There is, however, an important caveat regarding their fitting formula. Their model successfully reproduces the upturn of the mass--concentration relation at high redshift ($z \gtrsim 2$) for both the WMAP7 and Planck cosmologies.
However, for the Planck cosmology, numerical simulations show that the concentration becomes nearly flat toward the highest halo masses, and no significant upturn is present up to the maximum halo mass at the redshifts $z=0$, $0.5$, and $1$. In contrast, for the WMAP7 cosmology, no upturn is found at $z=0$ or $0.5$, whereas a clear upturn appears at $z=1$, in good agreement with the simulation results.

For numerical stability, we reparameterize the parameters as follows,
\begin{align}
    f_d&= f_{d,{\rm min}}+(f_{d,{\rm max}}-f_{d,{\rm min}})\frac{1}{1+e^{-\eta_{f_d}}}.
\end{align}
Here, $f_d$, which represents $a_0$, $b_0$, $a_1$, $b_1$, and $c_\alpha$, respectively, is parameterized by the sigmoid function, $S({\eta})$, to restrict the parameter region of $[f_{d,{\rm min}}, f_{d,{\rm max}}]$. We set $a_{0,{\rm min}}=b_{0,{\rm min}}=a_{1,{\rm min}}=b_{1,{\rm min}}=1$, $a_{0,{\rm max}}=b_{0,{\rm max}}=a_{1,{\rm max}}=b_{1,{\rm max}}=5$, $c_{\alpha,{\rm min}}=0$, and $c_{\alpha,{\rm max}}=1$, respectively. 
In summary, the parameter set is given by ${\bm \theta}=\{\eta_{a_0}, \eta_{b_0}, \eta_{a_1}, \eta_{b_1}, \eta_c, \kappa\}$. 
We fix $\kappa=0.41$ and $\eta_c=-\ln((1/0.21)-1)$ following \citet{2019ApJ...871..168D} because they cannot be constrained.

\subsection{Parent populations}

The parent X-ray cluster catalogs are constructed using flux-limited selection, sometimes with additional criteria based on the extended nature of the X-ray emission. However, the WL analysis imposes further selection criteria, such as the removal of clusters overlapping with other catalogs and the availability of a sufficiently high-quality WL shape catalog that is less affected by stellar masks and other optical observational systematics. 
Therefore, it is difficult to quantify the purity, completeness, and Eddington bias for each subsample. In particular, it is difficult to model the observed distribution of true masses for the WL-targeted clusters using the halo mass function, because the number of low-mass systems rapidly decreases near the observational detection threshold.

On the other hand, this problem can be treated more straightforwardly by introducing latent parameters that describe the parent true-mass distribution and fitting them simultaneously with the other model parameters \citep[e.g.][]{2016MNRAS.455.2149S,2022PASJ...74..175A,2025A&A...700A..46O}. This approach allows us to account for the underlying mass distribution of the parent cluster population without requiring a fully characterized selection function for the WL-targeted samples.

The parent distribution of $X$ is modeled as a Gaussian, $\mathcal{N}(\mu_X(z),\sigma_X(z))$, with redshift-dependent hyper-parameters $\bm{\psi}=(\mu_X(z),\sigma_X(z))$. We parameterize this dependence as $\bm{\psi}(z)=\bm{\alpha}_{\psi} + \sum_k^{N_\psi} \bm{\gamma}_{\psi,k} F(z)^k$. For the parent population, we adopt the median redshift, $z_p=0.265$, as the pivot redshift in $F(z)$. The polynomial orders, $N_\psi$, used in this parameterization are listed in Table \ref{tab:orderNpsi}. The appropriate order can be determined by examining whether increasing the order leads to a better-constrained model.

\begin{table}[]
    \centering
    \begin{tabular}{c|cc}
      Sample           &  $\mu_X$ & $\sigma_X$ \\ 
                \hline
      eRASS1   &  1 & 1\\
      XXL & 2 & 1 \\
      LoCuSS & 0 & 0 \\
      MCXC  &  1 & 0 
    \end{tabular}
    \flushleft
    \caption{The polynomial orders used to describe the redshift dependence of the underlying true mass distribution.}
    \label{tab:orderNpsi}
\end{table}

\subsection{Priors}

We use the flat prior of the normalization parameter, $\alpha$, in $[-50,50]$. For the parameters describing the mass and redshift dependence, $\beta$ and $\gamma$, we adopt Student-$t$ priors with one degree of freedom in order to obtain a prior that is uniform in angle, as follows
\begin{align}
    p(\beta)=\frac{1}{\pi}\frac{1}{1+\beta^2}.
\end{align}
When we introduce the reparameterized variable, $\xi$, to allow the slope to take either positive or negative values, the prior is transformed consistently by including the corresponding Jacobian factor. Specifically, if the physical slope is written as $\beta=f(\xi)$ and the desired prior on the slope is the Student-$t$ prior with one degree of freedom, then the prior on $\xi$ is given by
\begin{align}
    p(\xi)=p\left(\beta \right)\left|\frac{d\beta}{d\xi}\right|.
\end{align}
Equivalently, this corresponds to adding the Jacobian contribution to the log-prior,
\begin{align}
   \ln p(\xi)=-\ln\pi-\ln\left[1+{\ln(1+e^\xi)}^2\right]-\ln(1+e^{-\xi}).
\end{align}
The last term is the logarithm of the sigmoid function, i.e. $+\ln S(\xi)$. 

In an analogous way, the prior for the reparameterized variable $\eta$ is given by
\begin{align}
    \ln p(\eta)=\ln S(\eta)+\ln(1-S(\eta)),
\end{align}
which yields a flat prior in the original parameter. 

The reparameterized forms for the intrinsic scatter and intrinsic coefficient are
\begin{align}
    \ln p(\ln \sigma)&=\ln \sigma, \\
    \ln p(u) &= \ln\left(1-\tanh^2u\right).
\end{align}

We also treat the WL mass bias parameters as free parameters,  under the assumption that their uncertainty can be described by a multivariate Gaussian prior:
\begin{align}
    p(\bm{\theta}_{\rm WL})
    = \mathcal{N}\!\left(\bm{\theta}_{\rm WL} \mid \bm{\theta}_{\rm WL}, \bm{\Sigma}_{\rm WL}\right),
\end{align}
where $(\bm{\theta}_{\rm WL})$ and $(\bm{\Sigma}_{\rm WL})$ denote the mean vector and covariance matrix based on the mock analysis (Table \ref{tab:WLcalib}), respectively. In our analysis, we employ the Affine reparameterization described in Appendix \ref{app:white}.

\subsection{Hamiltonian Monte Carlo Method}

We sample the posterior distribution using the Hamiltonian Monte Carlo (HMC) method, implemented with the Python package \texttt{BlackJAX} in the \texttt{JAX} framework. HMC is more efficient than standard random-walk Markov chain Monte Carlo methods, especially in high-dimensional parameter spaces, because it uses gradient information of the log posterior to propose distant states while maintaining a high acceptance probability. This reduces random-walk behavior and improves the exploration of correlated parameters.

The computations are accelerated using a graphics processing unit (GPU), which is particularly useful for evaluating the likelihood and its gradients repeatedly during the HMC sampling. We assess the convergence and reliability of the chains using standard diagnostics, including the effective sample size (ESS), the potential scale reduction factor ($\hat{R}$), the acceptance fraction, and the number of divergent transitions related to energy conservation. We also monitor the behavior of the Hamiltonian energy to check that the numerical integration is sufficiently accurate. These diagnostics confirm that the HMC sampler provides stable and well-converged posterior samples.

Although HMC substantially reduces the random-walk behavior of standard MCMC methods, its sampling efficiency can still degrade when the posterior distribution has a strongly curved or banana-shaped geometry. Such a geometry indicates that the local correlation structure of the parameters changes across the posterior, making it difficult for a fixed mass matrix and numerical Hamiltonian integration to explore the distribution efficiently. To mitigate this problem, we internally adopted the affine transformation (Appendix \ref{app:white}).
We validated the correctness of the technical implementation in the code by comparing it with the case of the linear WL mass calibration \citep{2025A&A...700A..46O}, which can be fully described by an analytical formula.

If the free parameters cannot be constrained, energy conservation in the HMC method tends to be violated. In that case, we either fix these parameters or impose a Gaussian prior on them.

\section{Result and Discussions} \label{sec:result}

\subsection{WL mass versus redshift}

\begin{figure}
    \centering
    \includegraphics[width=\linewidth]{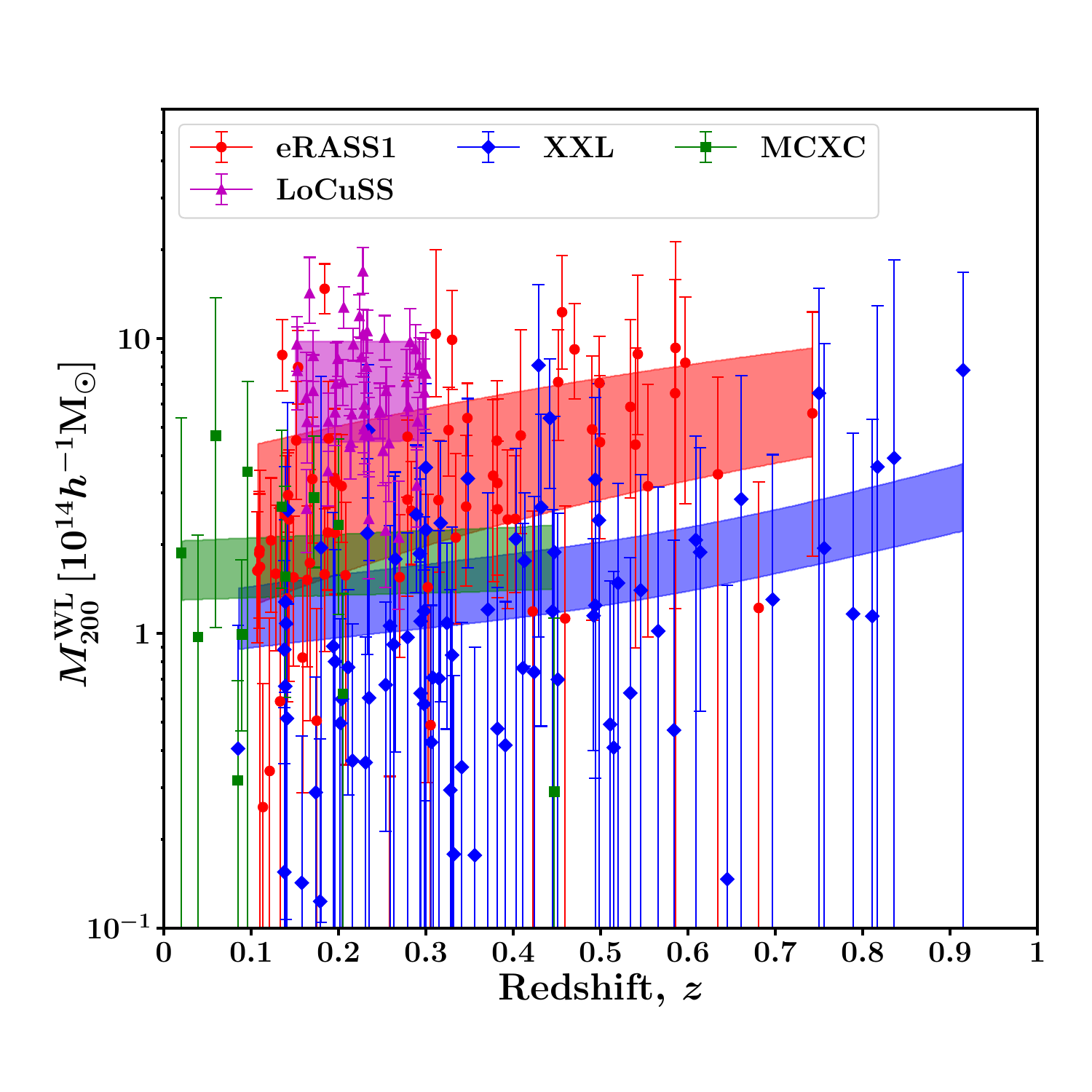} 
    \flushleft
    \caption{The WL masses versus redshift. The red circles, blue diamonds, magenta upward triangles, and green squares correspond to the eRASS1, XXL, LoCuSS, and MCXC samples, respectively. The transparent red, blue, magenta, and green areas indicate the respective $68.3\%$ intervals for the WL masses. {Alttext: one figure in color.}}
    \label{fig:z_vs_M200}
\end{figure}

Figure \ref{fig:z_vs_M200} presents the redshift distribution of the WL-derived masses for the 213 clusters. Overlaid on the figure are the $1\sigma$ confidence regions for the probability distribution of the true WL masses for each sample, calculated as
\begin{align}
    p({\tilde x})=\int dX\, p({\tilde x}|X, z, {\bm \theta}_{\rm WL})\,p(X|z,{\bm \psi}).
\end{align}
In this calculation, we adopt the SP model, although using alternative concentration models yields no significant changes in the results.
The LoCuSS sample probes the most massive systems in the range $0.15 \le z \le 0.3$. The MCXC sample, on the other hand, traces intermediate-mass systems at low redshift. The eRASS1 covers a higher mass range, whereas XXL extends to lower masses and higher redshifts.
Therefore, each sample provides complementary coverage of the mass and redshift ranges used in our Bayesian analysis.

\subsection{WL concentration and WL mass distribution}

\begin{figure}
    \centering
    \includegraphics[width=\linewidth]{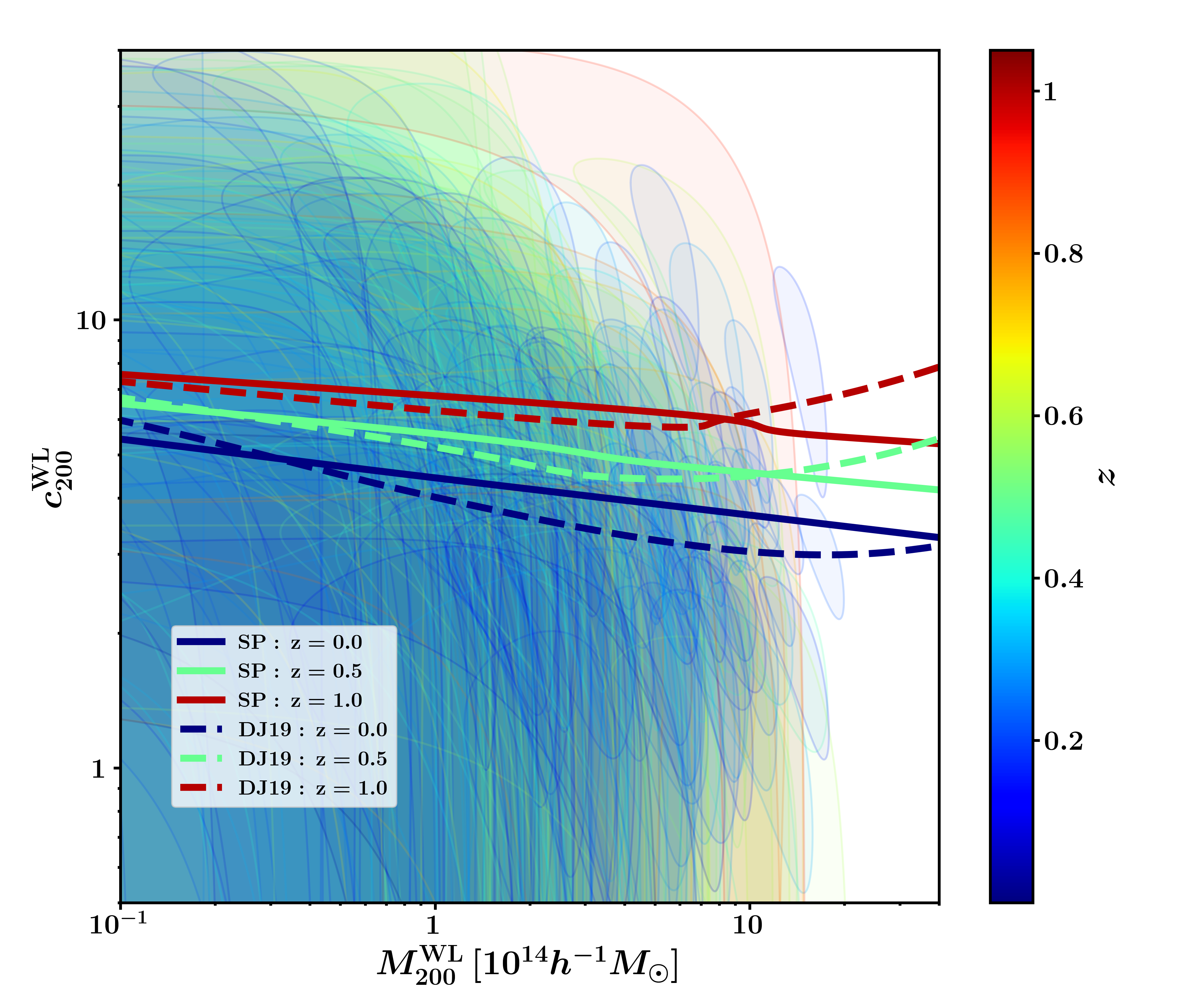} 
    \flushleft
    \caption{The translucent banana-shaped regions indicate the $1\sigma$ constraints on the WL mass and halo concentration. The color scale shows the cluster redshift. The solid and dashed curves indicate the WL normalizations for the SP and DJ19 models, respectively, where the redshifts $z=0$, $0.5$, and $1$ are shown in blue, light green, and red. {Alttext: one figure in color.}}
    \label{fig:MC_WL}
\end{figure}

The WL concentration and WL mass distributions are shown in Figure \ref{fig:MC_WL}. The clusters at higher redshifts are shown in red, those at lower redshifts are shown in blue, and clusters at intermediate redshifts are shown in green. It is found that the red region appears in the high-mass range, the green region in the intermediate-mass range, while the blue region is spread more uniformly. For comparison, we plot the baselines of the true WL mass--concentration relation calculated by 
\begin{align}
  p({\tilde y}|{\tilde x})=\int d^2 {\bm X} p({\tilde {\bm x}}|{\bm X}, z,{\bm \theta}_{\rm WL})p(Y|X,z,{\bm \theta},{\bm \theta}_{\rm cosmo}).
\end{align}
Here we adopt the SP and DJ19 models. The first thing to emphasize is that our WL data do not extend into the mass regime where the concentration upturn becomes pronounced, but instead lie in the transitional mass range. Consequently, the fitting results mainly capture the curvature of the baseline of the mass--concentration relation for low- and intermediate-mass clusters.


\subsection{Mass-concentration relation}
In the actual HMC analysis, we employ the reparameterized variables to describe the baseline of $\mu$. Because some of the scaling relations use different parameterizations, we transform the posterior samples back to the original parameters and then estimate them using a central bi-weight estimator \citep{1990AJ....100...32B} of the marginalized distributions for convenience. The uncertainties are computed as the interval extending a distance of $0.34135$ from the central bi-weight estimate\footnote{This is the central $68.27\%$ credible interval. We do not use the bi-weight scale estimator \citep{1990AJ....100...32B} as the error estimates.}, which we denote as the $1\sigma$ error. The inferred parameters are listed in Tables \ref{tab:SPB13}-\ref{tab:DJ19}. Whenever a Gaussian prior is used, this is indicated in the table captions. The baseline measurements and their corresponding uncertainties are shown in Figure \ref{fig:cmrelation}.

\subsubsection{SP model}

The result of the SP model is shown in Table \ref{tab:SPB13}.
The measurement uncertainties of the mass-dependent slope, $\beta$, and the redshift-dependent slope, $\gamma$, are reduced by about factors of 4, compared to those obtained for the eRASS cluster sample \citep{2025A&A...700A..46O}, thanks to a larger number of samples and the reparameterized method. 
We also succeeded in constraining the intrinsic scatter, whereas previous studies were able to place only an upper limit on it.

The inferred baseline (Figure \ref{fig:cmrelation}) is consistent, within $1\sigma$ uncertainty, with the numerical simulations \citep{2013ApJ...766...32B,2018ApJ...859...55C,2015ApJ...799..108D,2019ApJ...871..168D,2021MNRAS.506.4210I} over the redshift interval $0\le z\le 1$, with the exception of the upturn mass regime reported by \citet{2015ApJ...799..108D} and \citet{2019ApJ...871..168D}. The deviations of the upturn mass-concentration relations from the SP model reach the $2\sigma$ and $3\sigma$ levels at a mass of $10^{15}h^{-1}M_\odot$ for redshifts $z=0.5$ and $z=1.0$, respectively.
Since our sample does not adequately cover the mass range where the upturn is predicted, the comparison in this regime is merely an extrapolation beyond the mass range directly probed by our data.

\citet{2019MNRAS.486.4001R} employed the SP model and obtained $\beta=-0.12$ and $\gamma=0.16$, values that are consistent with our findings within the $1\sigma$ uncertainties.

\begin{table}[]
    \centering
    \begin{tabular}{c|cc}
     & SP & B13 \\
     \hline
$\alpha$ & $1.50_{-0.12}^{+0.14}$  & $1.64_{-0.20}^{+0.29}$  \\
$\beta$ & $-0.08_{-0.07}^{+0.05}$ &  $-0.25_{-0.25}^{+0.17}$  \\
$\gamma$ & $-0.32_{-0.40}^{+0.23}$  & $0.37_{-0.27}^{+0.50}$ \\
$\sigma_Y$ & $0.07_{-0.05}^{+0.07}$ & $0.07_{-0.05}^{+0.07}$  \\
    \end{tabular}
 \flushleft
    \caption{The inferred parameters of the SP (eq. \ref{eq:SP}) and B13 (eq. \ref{eq:B13}) models. }
    \label{tab:SPB13}
\end{table}

\subsubsection{B13 model}

The result of the B13 model is shown in Table \ref{tab:SPB13}. The baseline has a weak non-linear mass dependence on the concentration parameters (eq. \ref{eq:B13} and Figure \ref{fig:cmrelation}). The baseline obtained from the numerical simulation \citep{2013ApJ...766...32B} is in good agreement with our inferred baseline. Compared to the SP model, the inferred B13 model yields a lower concentration in the low-mass regime around $\sim 10^{13}h^{-1}M_\odot$, and this difference increases at higher redshifts.

\subsubsection{C18 model}

The parameter, $\ln b$, which controls the transition scale, enters the sigmoid function and is therefore difficult to constrain directly. We consequently impose a Gaussian prior $p(\ln b) = \mathcal{N}(6, 0.5)$ on this parameter and adopt a value close to the theoretical expectation, $\ln b = 6.09$. When the prior was removed, energy conservation in the HMC trajectories broke down, preventing the posterior from being sampled properly.
Similarly, the small-quantity parameter $\ln c_r$ is also poorly constrained, so we impose a Gaussian prior $p(\ln c_r) = \mathcal{N}(0, 1)$, where its center is close to the theoretical prediction $\ln c_r=-0.075$.
The results for the C18 model are summarized in Table \ref{tab:C18}. When we double the prior standard errors, the inferred baseline remains unchanged because of parameter degeneracy, although the baseline uncertainty increases at $M_{200}\simlt 3\times 10^{13}h^{-1}M_\odot$.
The inferred baseline closely matches the simulation outcome (Figure \ref{fig:cmrelation}); this similarity arises from the priors that were employed.
When we shift the centers of the Gaussian priors to $p(\ln b)={\mathcal N}(3,0.5)$ and $p(\ln c_r)={\mathcal N}(-1,1)$, the inferred parameters become $A=4.37_{-2.58}^{+5.70}$, $b=21.56_{-8.73}^{+13.57}$, $c_r=0.35_{-0.31}^{+1.31}$, and $\beta=-0.26_{-0.40}^{+0.19}$. However, due to parameter degeneracies, the baseline remains largely unchanged at $M_{200}\simgt 10^{14}h^{-1}M_\odot$ for $z=0$ and $M_{200}\simgt 10^{13}h^{-1}M_\odot$ for $z=1$, respectively.

\begin{table}[]
    \centering
    \begin{tabular}{c|c}
         &  C18 \\
         \hline
$A$ & $1.47_{-0.88}^{+1.97}$  \\
$b$ &  $459.51_{-189.02}^{+303.20}$ \\
$c_0$ & $2.82_{-2.54}^{+10.90}$  \\
$\beta$ &  $-0.12_{-0.18}^{+0.09}$ \\
$\sigma_Y$ & $0.07_{-0.05}^{+0.07}$ \\
    \end{tabular}
    \flushleft
    \caption{The inferred parameters of the C18 (eq. \ref{eq:C18}). We use Gaussian priors of  $p(\ln b)={\mathcal N}(6,0.5)$ and $p(\ln c_r)={\mathcal N}(0,1)$.}
    \label{tab:C18}
\end{table}

\subsubsection{DP model}

The result of the DP model is shown in Table \ref{tab:DP} and Figure \ref{fig:cmrelation}. Since we cannot constrain $\gamma_t$, which describes the redshift evolution associated with a change in the upturn mass range, we set $\gamma_t = 0$. Furthermore, we use a Gaussian prior of $p(\alpha_t)={\mathcal N}(\ln(10),\ln(10))$. The resulting mass scale associated with the upturn slope, $M_t=6.36_{-5.27}^{+13.56} \times 10^{14}h^{-1}M_\odot$, is only weakly constrained. 
The baseline uncertainty increases for $M_{200}\simgt 10^{15}h^{-1}M_\odot$ and $M_{200}\sim \mathcal{O}(10^{13}h^{-1}M_\odot)$ because a small sample of clusters exist within these mass ranges. Because the upturn slope is steep $\beta_2=0.51$, the inferred $\beta$ becomes larger so that the mass-concentration relation remains relatively flat over the mass range where most of the sample lies.  

\citet{2016MNRAS.457.4340K} showed a similar form of eq. \ref{eq:C18}. They reported that the mass scale of the upturn is $M_t=5.5\times10^{16} h^{-1}M_\odot$ at $z=0$ and $9\times 10^{14}h^{-1}M_\odot$ at $z=1$, with the upturn slope $\beta_2\sim 0.52$. 
Their mass scale of the upturn is the same as that of our analysis in that it cannot be constrained.

When we adopt a prior of $p(\alpha_t)={\mathcal N}(0,\ln(10))$, the inferred value of $M_t$ becomes $1.36_{-1.16}^{+4.26}\times 10^{14}h^{-1}M_\odot$, while the baseline results remain largely unchanged because of parameter degeneracies. Due to the statistical significance (Section \ref{subsec:stat}), we employ the former prior. 

The inferred baseline agrees with the theoretical models within a $1\sigma$ uncertainty. However, the mass dependence on the concentration parameter in the low-mass regime is entirely different from what is found in numerical simulations. To evaluate this discrepancy, we require a large sample of clusters within this mass range, which is left to be addressed in future work.

\begin{table}[]
    \centering
    \begin{tabular}{c|c}
     & DP \\
     \hline
$\alpha$ &  $1.43_{-0.43}^{+0.26}$\\
$\beta$&  $-0.34_{-0.39}^{+0.22}$  \\
$\gamma$& $-0.32_{-0.39}^{+0.23}$\\
$\alpha_t$ &  $1.96_{-1.64}^{+1.16}$ \\
$\gamma_t$ & (0) \\
$\beta_2$ &  $0.51_{-0.38}^{+0.80}$ \\
$\sigma_Y$ & $0.06_{-0.04}^{+0.07}$ \\
    \end{tabular} 
    \flushleft
    \caption{The inferred parameters of the DP (eq. \ref{eq:DP}). We use a Gaussian prior of $p(\alpha_t)={\mathcal N}(\ln(10),\ln(10))$. A parameter shown in parentheses is fixed. }
    \label{tab:DP}
\end{table}

\subsubsection{DK15 model}

The results of the DK15 model are presented in Table \ref{tab:DK15} and Figure \ref{fig:cmrelation}.
We cannot constrain the scale parameters of $\nu$ at which the slope changes, so we fix them to the median values predicted by the theoretical DK15 model. The results of the DK15 model are presented in Table \ref{tab:DK15} and Figure \ref{fig:cmrelation}. Because the slope parameters are treated as free, the uncertainties in the baseline become large at both the high- and low-mass ends. Within these uncertainties, the inferred baseline is consistent with the DK15 model.

\begin{table}[]
    \centering
    \begin{tabular}{c|c}
    & DK15\\
    \hline
    $\phi_0$ & $3.49_{-0.42}^{+0.50}$ \\\
    $\phi_1$ & $2.07_{-1.26}^{+2.40}$ \\\
    $\eta_0$ & (1.56)\\
    $\eta_1$ & (7.28) \\
    $\beta$ & $-0.99_{-0.55}^{+0.58}$\\
    $\beta_2$ & $0.79_{-0.62}^{+1.56}$\\
    $\kappa$ & (1)\\
     $\sigma_Y$ & $0.06_{-0.04}^{+0.07}$ \\
    \end{tabular}
    \flushleft
    \caption{The inferred parameters of the DK15 (eq. \ref{eq:DK15}). We use a Gaussian prior of $p(\ln \phi_0)={\mathcal N}(2,1)$ and $p(\ln \phi_r)={\mathcal N}(-1.5,1)$. A parameter shown in parentheses is fixed.}
    \label{tab:DK15}
\end{table}

\subsubsection{DJ19 and I21 models}

\begin{figure*}
    \centering
    \includegraphics[width=\linewidth]{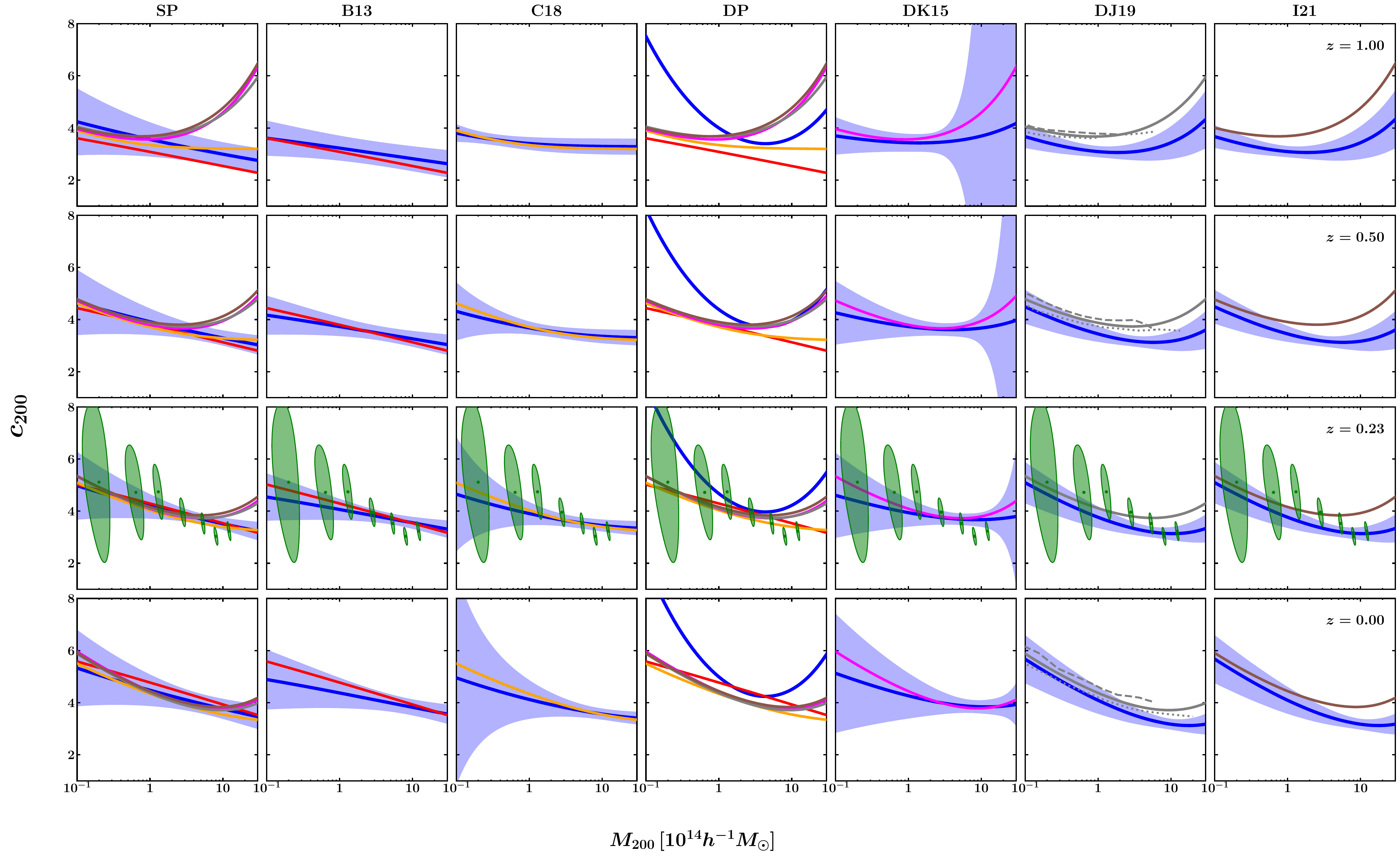}
    \flushleft
    \caption{The inferred mass-concentration relation. From left to right, the panels correspond to SP, B13, C18, DP, DK15, DJ19, and I21. From top to bottom, the rows show the results at $z=1$,  $z=0.5$, $z=0.23$ (which is close to the redshift of the stacked quantities), and $z=0$, respectively. The blue curves and their shaded regions indicate the inferred baseline relation and its $1\sigma$ uncertainty. The red, orange, magenta, gray and brown lines represent the predictions of the B13, C18, DK15, DJ19, and I21 models, respectively.  The green ellipses in the 3rd row mark the stacked quantities. In the DJ19 column, the dashed and dotted curves represent the Planck- and WMAP-based simulation results obtained from \citep{2019ApJ...871..168D}, respectively.  {Alttext: A figure consisting of seven columns and four rows of panels.}}
    \label{fig:cmrelation}
\end{figure*}

The results of the DJ19 model are presented in Table \ref{tab:DJ19} and Figure \ref{fig:cmrelation}.
We cannot constrain $c_\alpha$ which is the degree of the effective exponent of linear growth on the normalization, and we thus fixed it to the median value of the theoretical DJ19 model.  Since I21 also used the same formula, we plot the I21 model in Figure \ref{fig:cmrelation}. The mass-dependent slope of the concentration parameter is fixed by the terms $\nu^{-1}$ and $\nu$. Therefore, the baselines exhibit lower uncertainty than the DP and DK15 models.
The concentrations predicted by the DJ19 and I21 models are higher than the inferred values at $M_{200}\gtrsim 10^{15}h^{-1}M_\odot$ for $z=0$ and at $M_{200}\gtrsim 10^{14}h^{-1}M_\odot$ for $z=1$, with discrepancies exceeding the $2\sigma$ level.

At the redshifts covered by our sample, the slope of the upturn appears to be shallower than that predicted by the model. Instead, the upturn term acts mainly to flatten the concentration-mass relation at the high-mass end, causing the concentration to approach an approximately constant value with increasing mass. This behavior is consistent with the prediction of C18 and \citet{2016MNRAS.457.4340K}.

An important caveat is that, although the upturn feature is clearly visible as a function of mass at $z\ge 2$ \citep{2019ApJ...871..168D}, the simulations at $z\le 1$ do not necessarily cover the mass range in which the upturn is expected to appear. In Figure \ref{fig:cmrelation}, we therefore overplot the Planck- and WMAP-based simulation results taken from Figure 3 of \citet{2019ApJ...871..168D} as dashed and dotted curves, respectively. The absence of a curve indicates a mass range not covered by the corresponding simulations.

Within the mass ranges covered by the simulations, these curves are closely similar to the relation inferred from the data. Our sample likewise does not sufficiently cover the mass regime in which the upturn is expected. Over the mass ranges jointly probed by the simulations and our observations, the inferred and simulated relations exhibit nearly the same shape, although the minimum normalization inferred from the observations is slightly lower.

\begin{table}[]
    \centering
    \begin{tabular}{c|c}
    & DJ19\\
    \hline
    $a_0$ & $2.07_{-0.63}^{+0.95}$\\
     $a_1$  &$2.51_{-1.04}^{+1.52}$  \\
     $b_0$ & $3.27_{-1.11}^{+1.08}$  \\
     $b_1$ & $3.29_{-1.27}^{+1.14}$ \\\
     $c_\alpha$ & $(0.21)$ \\
    $\kappa$ & $(0.41)$ \\
     $\sigma_Y$ & $0.06_{-0.05}^{+0.07}$ 
    \end{tabular} 
    \flushleft
    \caption{The inferred parameters of the DJ19 model (eq. \ref{eq:DJ19}). Parameters shown in parentheses are fixed. }
    \label{tab:DJ19}
\end{table}

\subsubsection{Comparisons with other models}

We did not perform Bayesian analysis using the other models of the numerical simulations because of issues such as numerical instabilities and the lack of continuous redshift coverage. Instead, we here discuss comparisons with other representative studies.

The baseline relation of \citet{2014MNRAS.441.3359D} is higher than that of the SP model at $z=0$ at the $1\sigma$ level, while it is in good agreement with the SP model at $z=1$. \citet{2016MNRAS.457.4340K} predict a mass-dependent slope that is consistent with that of C18 over $0 \le z \le 1$, although their normalization is higher than that inferred for the C18 model at the $\simlt 1\sigma$ level. \citet{2016MNRAS.460.1214L} reported a steeper mass dependence and a higher normalization; however, their results remain consistent with the SP model within the $1\sigma$ uncertainty. The baselines of \citet{2012MNRAS.423.3018P} are higher than those of the SP and DJ19 models at the $2\sigma$ level for $z=0$ and at the $\simgt 5\sigma$ level for $z=1$, indicating statistically significant discrepancies from our results for all considered models.

\subsection{Weighted mean data points in the true mass-concentration plane}

For visualization purposes, we compute weighted means of the data points in the true mass-concentration plane of $(X,Y)$. Due to the chain rule, we convert from ${\tilde {\bm x}}$ to ${\bm x}$, and from ${\bm \Sigma}_{\rm WL}$ to ${\bm \Sigma_{\rm err}}+{\bm \Sigma}_{\rm WL}$.
We infer the true quantities, ${\bm X}$, from the WL observables, ${\bm x}$, by minimizing 
\[
\frac{1}{2}({\bm x}-{\bm f}_{\rm WL}({\bm X}))^{T}({\bm \Sigma_{\rm err}}+{\bm \Sigma}_{\rm WL})^{-1}({\bm x}-{\bm f}_{\rm WL}({\bm X})).
\]
In this step, the best-fit parameters of the WL mass calibration are held fixed. We then compute the covariance matrix in the ${\bm X}$ plane as
\begin{align}
    {\bm J} &= \frac{\partial {\bm f}_{\rm WL}({\bm X})}{\partial {\bm X}}, \\
    {\bm \Sigma}_{\bm X} &= \left({\bm J}^T({\bm \Sigma_{\rm err}}+{\bm \Sigma}_{\rm WL})^{-1}{\bm J}\right)^{-1}.
\end{align}
Using these, we evaluate the weighted mean quantities via
\begin{align}
    {\bm X} &= \left( \sum_i {\bm \Sigma}_{{\bm X},i}^{-1}\right)^{-1}\sum_i {\bm \Sigma}_{{\bm X},i}^{-1}{\bm X}_i.
\end{align}
The subsamples are constructed by binning in mass rather than in redshift, so as to distinguish between different curvatures of the model baselines. The average redshifts are estimated by a weight of $1/\sqrt{{\rm det}{{\bm \Sigma}_{\bm X}}}$. 
The results are shown in the 3rd row of Figure \ref{fig:cmrelation}.

It is not possible to distinguish from the weighted mean data point in the 3rd row of Figure \ref{fig:cmrelation} with statistical significance whether the data points are distributed in a straight line or in a curved pattern.

\subsection{Statistical significance} \label{subsec:stat}

To evaluate the predictive performance of different models, we compute the Widely Applicable Information Criterion \citep[WAIC;][]{Watanabe2010} and the Pareto-smoothed importance sampling leave-one-out information criterion \citep[PSIS-LOOIC;][]{Vehtari2017} from the pointwise log-likelihood. Both criteria estimate the expected predictive accuracy for unseen data while penalizing model complexity. In Bayesian model comparison, we adopt PSIS-LOOIC as our primary model-selection criterion because it provides a more reliable approximation to exact leave-one-out cross-validation and offers diagnostic information through the Pareto shape parameter ($\hat{k}$). WAIC is reported as a complementary measure. We use the Python package \texttt{ArviZ} for the computation. 

We use the linear mass-concentration relation (SP) model as a reference to evaluate the differences in LOOIC and WAIC, specifically, $\Delta{\rm LOOIC} = {\rm LOOIC}_{\rm model} - {\rm LOOIC}_{\rm SP}$ and $\Delta{\rm WAIC} = {\rm WAIC}_{\rm model} - {\rm WAIC}_{\rm SP}$.

The results are summarized in Table \ref{tab:looic}. We find no discrepancy above the $4\sigma$ threshold; however, the C18 and B13 model are favored over the SP model at approximately the $2\sigma$ and $1\sigma$ levels, respectively. The main difference between C18 and C13 is a sign of the second derivative of $\ln c_{200}$ with respect to $\ln M_{200}$; 
\begin{eqnarray}
    \frac{d^2\ln c_{200}}{d(\ln M_{200})^2}\Bigl|_{\rm B13}<0, \quad    \frac{d^2\ln c_{200}}{d(\ln M_{200})^2}\Bigl|_{\rm C18}>0.
\end{eqnarray}
The B13 model indicates that its logarithmic slope becomes progressively more negative toward higher masses. In contrast, the C18 is concave upward, indicating that the mass dependence gradually flattens toward the high-mass end. When we evaluate $\Delta(\mathrm{LOOIC})$ for C18 relative to B13, the C18 model is favored at the $1.7\sigma$ significance level. 
The DP, DK15, and DJ19 models all have a positive second derivative, $d^2\ln c_{200}/d(\ln M_{200})^2>0$, as does the C18 model.
When we incorporate the upturn components of the DP, DK15, and DJ19 models, the level of significance decreases. 
The difference of $\Delta$LOOIC of the C18 model relative to DP, DK15, and DJ19 models is better at the $\sim 0.5\sigma$, $\sim 1\sigma$, and $\sim 1\sigma$ levels, respectively. 
Even when we change the prior on the parameters of the C18 model, the result does not change. 
Therefore, our result captures a trend of gradual flattening of the mass-dependent slope. 
This feature is also present in the DJ19 simulation data (Figure \ref{fig:cmrelation}), though the analytical model extrapolates beyond the mass range covered by the simulations.

To place tighter constraints on the mass-concentration parameters, Stage-IV surveys such as Euclid \citep{Laureijs2011}, Rubin \citep{Ivezic2019}, and Roman \citep{Spergel2015} are essential, particularly when combined with the developed Bayesian methodology. This study considers a heterogeneous sample comprising several tens of clusters, for which we assume a Gaussian distribution for the true cluster masses and simultaneously infer the corresponding distribution parameters. Once a well-defined sample containing on the order of thousands of clusters becomes available, the parent population should instead be modeled using the full mass function, with a sharp low-mass cutoff imposed by the survey selection function. In that regime, a more sophisticated numerical integration scheme will be required, and the development and implementation of such a method are deferred to future work. The WL mass calibrations based on numerical simulations are also important. The issues of the mass-concentration relation could be addressed using both large observational and simulation data sets. In particular, if the data cover the upturn mass range, it would be helpful to distinguish between the upturn model and the model in which the concentration asymptotically approaches a constant value at high masses.

\begin{table}[]
    \centering
    \begin{tabular}{c|cc}
         &  $\Delta$LOOIC & $\Delta$WAIC\\
         \hline
      SP   &  0 &  0\\
      B13  & $-0.51\pm0.38 \,(1.4)$ & $-0.36\pm0.35\,(1.0)$ \\
      C18  & $-1.32\pm0.55 \,(2.4)$ & $-1.13\pm0.52\,(2.2)$ \\
      DP & $-0.55\pm1.39 \,(0.4)$ & $-0.41\pm1.39\,(0.3)$\\
      DK15 & $+1.39\pm2.86\,(0.5)$ & $+0.95\pm2.57\,(0.4)$\\ 
      DJ19 & $+1.86\pm2.92\,(0.6)$ & $+0.75\pm2.90\,(0.3)$ \\
    \end{tabular}
    \flushleft
    \caption{$\Delta$(LOOIC) and $\Delta$(WAIC). Parameters shown in parentheses present the signal-to-noise ratios of the model difference.}
    \label{tab:looic}
\end{table}

\section{Summary} \label{sec:summary}

We developed a generalized Bayesian framework to constrain a wide variety of mass-concentration relations motivated by both theoretical and observational studies. The framework incorporates the intrinsic scatter in concentration, nonlinear WL calibration relations, the WL measurement-error covariance, and the underlying parent mass distribution. The computation involves a two-dimensional numerical integration over the plane defined by the true mass and true concentration, and can therefore accommodate non-linear mass-concentration relations derived from numerical simulations. We considered a conventional single power-law model, an empirical double power-law model, and theoretical models calibrated against numerical simulations by B13 \citep{2013ApJ...766...32B}, C18 \citep{2018ApJ...859...55C}, DK15 \citep{2015ApJ...799..108D}, and DJ19 \citep{2019ApJ...871..168D}. We employed the GPU-based Hamiltonian Monte Carlo.

We compiled 213 WL-measurable, X-ray-selected clusters drawn from the eRASS1, XXL, LoCuSS, and MCXC catalogs at $z\simlt 0.9$.
For every subsample, we incorporated its individual WL calibration relation along with its corresponding underlying population distribution.
Because the mass and redshift ranges of our cluster sample do not extend sufficiently into the upturn regime predicted by numerical simulations, we distinguish among the models primarily through the curvature of their mass-concentration relations. 
Regardless of the model chosen, the resulting mass-concentration relation, along with its evolution with redshift, matches well with the findings of recent numerical simulations.
We find no statistically significant differences among the models at or above the $4\sigma$ level. However, the data show the C18 model is preferred at the $\sim2\sigma$ level, compared to the SP and B13 models. The key feature of the C18 model has a monotonically decreasing model with a positive second derivative of logarithmic concentration with respect to logarithmic mass, $d^2\ln c_{200}/d(\ln M_{200})^2>0$. This behavior agrees with the simulation results of \citet{2019ApJ...871..168D}, rather than with the analytical model, over the redshift interval $z \le 1$. The C18 model is also preferable to the double power-law models (DP, DJ19, and I21) at the $\sim 0.5-1\sigma$ level.

\section*{Acknowledgements}

The Hyper Suprime-Cam (HSC) collaboration includes the astronomical communities of Japan and Taiwan, and Princeton University.  The HSC instrumentation and software were developed by the National Astronomical Observatory of Japan (NAOJ), the Kavli Institute for the Physics and Mathematics of the Universe (Kavli IPMU), the University of Tokyo, the High Energy Accelerator Research Organization (KEK), the Academia Sinica Institute for Astronomy and Astrophysics in Taiwan (ASIAA), and Princeton University.  Funding was contributed by the FIRST program from the Japanese Cabinet Office, the Ministry of Education, Culture, Sports, Science and Technology (MEXT), the Japan Society for the Promotion of Science (JSPS), Japan Science and Technology Agency  (JST), the Toray Science  Foundation, NAOJ, Kavli IPMU, KEK, ASIAA, and Princeton University.

This paper makes use of software developed for the Large Synoptic Survey Telescope. We thank the LSST Project for making their code available as free software at  http://dm.lsst.org

This paper is based [in part] on data collected at the Subaru Telescope and retrieved from the HSC data archive system, which is operated by Subaru Telescope and Astronomy Data Center (ADC) at NAOJ. Data analysis was in part carried out with the cooperation of Center for Computational Astrophysics (CfCA), NAOJ. We are honored and grateful for the opportunity of observing the Universe from Maunakea, which has the cultural, historical and natural significance in Hawaii.

The Pan-STARRS1 Surveys (PS1) and the PS1 public science archive have been made possible through contributions by the Institute for Astronomy, the University of Hawaii, the Pan-STARRS Project Office, the Max Planck Society and its participating institutes, the Max Planck Institute for Astronomy, Heidelberg, and the Max Planck Institute for Extraterrestrial Physics, Garching, The Johns Hopkins University, Durham University, the University of Edinburgh, the Queen's University Belfast, the Harvard-Smithsonian Center for Astrophysics, the Las Cumbres Observatory Global Telescope Network Incorporated, the National Central University of Taiwan, the Space Telescope Science Institute, the National Aeronautics and Space Administration under grant No. NNX08AR22G issued through the Planetary Science Division of the NASA Science Mission Directorate, the National Science Foundation grant No. AST-1238877, the University of Maryland, Eotvos Lorand University (ELTE), the Los Alamos National Laboratory, and the Gordon and Betty Moore Foundation.

This work is based on data from eROSITA, the soft X-ray instrument aboard SRG, a joint Russian-German science mission supported by the Russian Space Agency (Roskosmos), in the interests of the Russian Academy of Sciences represented by its Space Research Institute (IKI), and the Deutsches Zentrum f\"{u}r Luft- und Raumfahrt (DLR). The SRG spacecraft was built by Lavochkin Association (NPOL) and its subcontractors, and is operated by NPOL with support from the Max Planck Institute for Extraterrestrial Physics (MPE). The development and construction of the eROSITA X-ray instrument was led by MPE, with contributions from the Dr. Karl Remeis Observatory Bamberg \& ECAP (FAU Erlangen-Nuernberg), the University of Hamburg Observatory, the Leibniz Institute for Astrophysics Potsdam (AIP), and the Institute for Astronomy and Astrophysics of the University of T\"ubingen, with the support of DLR and the Max Planck Society. The Argelander Institute for Astronomy of the University of Bonn and the Ludwig Maximilians Universit\"{a}t Munich also participated in the science preparation for eROSITA.

XXL is an international project based around an XMM Very Large Programme surveying two 25 deg$^2$ extragalactic fields at a depth of $\sim6 \times 10^{-15}\, {\rm erg\,cm^{-2}\,s^{-1}}$ in the [0.5-2] keV band for point-like sources. The XXL website is http://irfu.cea.fr/xxl. Multi-band information and spectroscopic follow-up of the X-ray sources are obtained through a number of survey programmes, summarised at http://xxlmultiwave.pbworks.com/.

The author gratefully thanks the HSC-SSP, eROSITA-HSC, and XXL-HSC collaborations. This work was supported by JSPS KAKENHI grant number JP25K07368.

The article processing charge was covered by the prize awarded through the PASJ Excellent Paper Award 2026 for the work "HSC-XXL: Baryon budget of the 136 XXL groups and clusters" by \citet{2022PASJ...74..175A}. The author extends heartfelt thanks to all co-authors.


\appendix

\section{Importance Sampling}

As for eq. (\ref{eq:likelihood}), double numerical integrations in the $X$ and $Y$ planes are computed by importance sampling.
We consider an integral involving a prior density $p(X,Y)$,
\begin{equation}
I \;=\; \int f(X,Y)\, p(X,Y)\, dX\, dY ,
\end{equation}
where $f(X,Y)$ typically represents a likelihood function. 
Given that we take a Gaussian distribution as the underlying distribution for both the X and Y axes, 
we generate quasi-random samples using a Sobol sequence transformed to a standard Gaussian distribution, $\mathcal{N}(0,1)$, 
where the Sobol sequence provides a low-discrepancy sampling of the unit hypercube before the Gaussian transformation.

For an arbitrary proposal distribution $q(X,Y)$, 
we rewrite
\begin{align}
I &= \int f(X,Y)\, p(X,Y)\, dX\, dY \\
&= \int f(X,Y)\, \frac{p(X,Y)}{q(X,Y)}\, q(X,Y)\, dX\, dY \\
&= {\rm E}_{q}\!\left[ f(X,Y)\,\frac{p(X,Y)}{q(X,Y)} \right].
\end{align}

The efficiency of importance sampling depends critically on the choice of the proposal distribution. In the ideal case, the variance-minimizing proposal is
\begin{equation}
q_{\rm opt}(X,Y)\propto f(X,Y)\,p(X,Y),
\end{equation}
for which the importance weights become constant. In practice, however, this optimal distribution cannot be sampled directly because it is known only up to an unknown normalization constant. We therefore construct a Gaussian approximation to the posterior distribution,
\begin{equation}
p(X,Y\mid{\bm x})
\propto
f(X,Y)\,p(X,Y),
\end{equation}
using a local linearization of the model around the posterior mode, and adopting this approximation as the proposal distribution $q(X,Y)$.

The practical benefit of this posterior-based proposal is that the samples are concentrated near the region that contributes most to the integral. Consequently, sharply peaked integrands can be evaluated more accurately with a given number of Sobol samples, resulting in more stable and efficient numerical integration.

\section{Affine reparameterization in Hamiltonian Monte Carlo} \label{app:white}

To improve the geometry of the target distribution of the HMC method, we sample transformed parameters instead of directly sampling the original physical parameters \citep[e.g.][]{2013arXiv1312.0906B}. Let ${\bm \theta}$ denote a subset of the original parameters. We introduce transformed parameters (${\bm u}$) through an affine transformation,
\begin{align}
{\bm \theta}={\bm \theta}_{0}+{\bm L}\,{\bm u},
\end{align}
where ${\bm \theta}_{0}$ is a reference parameter vector and ${\bm L}$ is a fixed, non-singular scaling matrix. Equivalently,
\begin{align}
{\bm u}={\bm L}^{-1}\left({\bm \theta}-{\bm \theta}_{0}\right).
\end{align}
Here, ${\bm u}$ is the parameter vector sampled by HMC, while the likelihood and prior are evaluated after transforming back to the physical parameters ${\bm \theta}$. The log posterior density in the transformed space is
\begin{align}
\ln p({\bm u}\mid {\bm x})=\ln p({\bm \theta} \mid {\bm x})+|\ln {\rm det}{\bm L}|. 
\end{align}

Since ${\bm L}$ is fixed, the Jacobian term is constant and does not affect the HMC trajectories, posterior gradients, or relative probabilities between samples. Thus, it can be omitted in the sampling. This affine reparameterization efficiently explores the parameter space while preserving the stability of HMC.

When a Gaussian prior $\mathcal{N}({\bm \mu}_{\rm prior}, {\bm C}_{\rm prior})$, is imposed on ${\theta}$, for instance, ${\bm \theta}_{\rm WL}$ and ${\bm \psi}$ in this paper, the affine transformation is given by
\begin{align}
{\bm \theta}_{0}={\bm \mu}_{\rm prior},\qquad
{\bm C}_{\rm prior}= {\bm L}_{\rm prior}{\bm L}_{\rm prior}^{\rm T},
\end{align}
where ${\bm L}_{\rm prior}$ is the lower triangular Cholesky factor. We then set 
\begin{align}
{\bm \theta}={\bm \mu}_{\rm prior}+{\bm L}_{\rm prior}\boldsymbol{u}.
\end{align}
The transformed variables, ${\bm u}$, follow a standard multivariate normal distribution. Equivalently, the prior is written by
\begin{align}
\ln p({\bm u})=-\frac{1}{2}{\bm u}^{\rm T}{\bm u}.
\end{align}
Gaussian prior whitening removes the scale differences and correlations encoded in the prior covariance matrix. In our \texttt{JAX/BlackJAX} implementation, HMC is performed in the transformed variables ${\bm u}$, while the likelihood is evaluated in the physical parameter space after applying the inverse transformation to obtain ${\bm \theta}$.  This transformation removes the scale differences and correlations encoded in the prior covariance matrix, making the local geometry of the target distribution closer to isotropic. As a result, HMC can typically use larger integration step sizes and explore the parameter space more efficiently.

\clearpage
\makeatletter
  \@in@appendixfalse
  \let\appendix\relax
  \let\appendixname\relax
  \let\appendixtitle\relax
  \def\@sectitle@prefix{}%
  \everypar{}%
\makeatother

\bibliographystyle{apj}
\bibliography{refs}

\end{document}